\documentclass[showpacs,preprint,preprintnumbers,amsmath,amssymb,endfloats]{revtex4}

\usepackage{amsmath,amssymb}
\usepackage{amsfonts}
\usepackage{epsfig,subfigure}
\usepackage{graphicx}
 
\def\br{{\bf r}}
\def\bk{{\bf k}}
\def\sp{{s^\prime}}

\newcommand{\expvdt}[1]{\left( #1 \right)_{\mathrm{D,T}}} 
\newcommand{\expvd}[1]{\left( #1 \right)_{\mathrm{D}}}

\newcommand{\expvt}[1]{\left \langle #1 \right \rangle}
\newcommand{\expv}[1]{\left \langle #1 \right \rangle}

\begin{document}

\title{Nonlinear optical spectroscopy of single, few, and many molecules; nonequilibrium Green's function QED approach} 
\author{Christoph A. Marx, Upendra Harbola and Shaul Mukamel}
\affiliation{Department of Chemistry, University of California, Irvine, CA 92697}
\date{\today}
\begin{abstract}
Nonlinear optical signals from an assembly of $N$ noninteracting particles consist of an incoherent and a coherent component, whose magnitudes scale $\sim N$ and $\sim N(N-1)$, respectively. A unified microscopic description of both types of signals is developed using a quantum electrodynamical (QED) treatment of the optical fields. Closed nonequilibrium Green's function expressions are derived that incorporate both stimulated and spontaneous processes. General $(n+1)$-wave mixing experiments are discussed as an example of spontaneously generated signals. When performed on a single particle, such signals cannot be expressed in terms of the $n$th order polarization, as predicted by the semiclassical theory. Stimulated processes are shown to be purely incoherent in nature. Within the QED framework, heterodyne-detected wave mixing signals are simply viewed as {\em{incoherent stimulated}} emission, whereas homodyne signals are generated by {\em{coherent spontaneous}} emission.
\end{abstract}

\maketitle

\section{introduction}

Nonlinear optical processes in bulk materials are traditionally classified as either coherent or incoherent \cite{Andrews}-\cite{Scully}. This classification refers to the relation of the macroscopic signal intensity of the sample to the nonlinear response of the individual molecules. Processes where the detected {\em{signal}} is obtained by simply adding up the contributions of the individual particles in the sample, are termed incoherent. Such signals hence scale as the number of particles $N$ in the interaction volume. With coherent processes, on the other hand, the contributions of the constituent particles add up on the level of {\em{amplitudes}} rather than intensities, and the detected signal is proportional to $N^2$. 

Coherent optical signals are usually calculated semiclassically by adopting a two step procedure \cite{Bloem65} - \cite{Scully}. First a (linear or nonlinear) polarization $P^{(n)}$ induced in the system by interactions with classical external fields is calculated microscopically. In the second step, $P^{(n)}$ is used as a source in the macroscopic Maxwell equations to generate the signal. Frequency-domain signals are given by the absolute square of an amplitude related to the susceptibilities $\chi ^{(n)}$ which, in turn, are proportional to $N$. This gives rise to the $\sim N^2$ dependence of coherent signals \cite{Muk115},\cite{Muk142}. Examples are Rayleigh scattering $(n=1)$, sum frequency generation $(n=2)$, four-wave mixing $(n=3)$. In contrast, signals such as Raman, fluorescence, two photon fluorescence etc. are incoherent and may not be calculated by the above semiclassical approach. Instead, they require a quantum description of the field to account for spontaneous emission. 

Treating both types of processes by a unified approach is of fundamental interest \cite{Andrews}, particularly when they coexist. For example, two photon fluorescence and second harmonic generation have been observed in the same sample \cite{Bidault}. There is a considerable current effort to perform nonlinear optical measurements on small samples and even single nano particles or molecules \cite{Orrit1}-\cite{muskens}. Recent experiments include 4-wave mixing spectroscopy of single semiconductor quantum dots \cite{Langbein} or nonlinear optical measurements on single electrons in F centers\cite{Lukin}. CARS microscopy \cite{Potma} is being extended to small interaction volumes. The traditional macroscopic formulation does not apply for small samples and a quantum electrodynamical (QED) formulation is called for.

A QED treatment is essential for predicting observables related to the quantum nature of the field. A powerful quantum master equation approach has been successfully used in quantum optics to describe fluctuations of the laser field, photon statistics, entanglement and squeezing \cite{Haroche}-\cite{Macovei}. In this paper we consider simple observables that are usually calculated by the semiclassical approach. However a quantum description of the fields allows a more precise definition of the signal, provides new insights and shows the limitations of the semiclassical approach. Our final results are expressed in terms of multipoint correlation functions of the material system. These can be evaluated for more complicated models than the Bloch-type equations ordinarily used in quantum optics. Details of molecular complexity, coupling to a solvent with arbitrary time scale and excitonic effects may be readily described by these expressions \cite{mukbook}.

The present approach is based on the many-body non-equilibrium Green's functions (NEGF) technique \cite{Negele} - \cite{Haug}. It provides a single microscopic definition of the signal that includes both spontaneous and stimulated emission and generates both incoherent and coherent signals. We shall use a superoperator in Liouville space representation that is commonly applied to study electrical conduction in quantum junctions \cite{Harbola_1}-\cite{Harbola_3}. In Sec. \ref{sec_optsig} we give a general definition of the signal (Eq. (\ref{eq_defsig})) and express it in terms of Green's functions of the system and the field. The formalism will be demonstrated by first considering the signal generated by a single molecule. Applications are made to two types of incoherent experiments, spontaneous light emission (SLE) (Sec. \ref{sec_sle}) and to pump-probe spectroscopy (Sec. \ref{sec_pumpprobe}). A purely diagrammatic derivation of the final expressions for the SLE signal using the technique of Keldysh Schwinger loops \cite{Mills} is developed. We present rules that directly translate the diagrams to partially-time-ordered expressions for the signal, both in the time- and in the frequency-domain. This representation, has been shown to be most useful for the study of frequency-domain experiments within the semiclassical approximation \cite{Muk566}. We extend it to quantum fields, and show that it yields more compact expressions than the fully-time-ordered double-sided Feynman diagrams. 

In Sec. \ref{sec_cohincoh}, we turn to the response of an assembly of noninteracting particles. The expressions of Sec. \ref{sec_optsig} are generalized by simply replacing the dipole operator of a single molecule by that of an assembly of particles. The signal is separated into an incoherent and a coherent part. The
latter, which is described by semiclassical optical susceptibilities, is dominant in macroscopic samples. However for small $N$ both have comparable magnitudes and coexist. The coherent signal does not exist in the limit of a single molecule. This indicates a breakdown of the semiclassical theory which does predict such a signal. Finally, in Sec. \ref{sec_het} we show that within QED heterodyne detection emerges as an incoherent stimulated emission
process in the detection mode. This is in contrast with semiclassical treatment where we first generate a coherent macroscopic signal, interfere it with a local oscillator and then obtain the signal. All of these steps are avoided by the microscopic QED definition given here. Sec. \ref{sec_concl} provides a summary of our results.

\section{Incoherent optical signals} \label{sec_optsig}

We first consider a single molecule driven by an optical field
$E(t)$. This is all we need for calculating incoherent signals. The extension to coherent signals from molecular ensembles will be done in Sec. \ref{sec_cohincoh}. 

The total Hamiltonian is given by
\begin{equation}
\hat{H}=\hat{H}_{0}+ \hat{H}_F+ \hat{H}_{int} ~\mbox{,}
\end{equation}
where $\hat{H}_{0}$ represents the free molecule, and
\begin{equation} \label{eq_fieldham}
\hat{H}_{F}=\sum_s\hbar \omega _{s}\hat{a}_{s}^{\dag}\hat{a}_{s} ~\mbox{,}
\end{equation}%
is the radiation field Hamiltonian. In Eq. (\ref{eq_fieldham}) $\hat{a}_s(\hat{a}_s^\dag)$ denotes the destruction (creation) operator for the $s$th field mode and $\hat{a}_{s}(\hat{a}_{s}^{\dagger })$ satisfy the boson commutation relation $[\hat{a}_{s},\hat{a}_{s^{\prime }}^{\dagger }]=\delta _{ss^{\prime }}$. For clarity, we shall omit the unit vector describing the field polarization of a particular mode, and treat the field as a scalar. The generalization to include the full tensorial expressions is straightforward.

We shall treat the emitted photon modes with frequency $\omega _{s}$ quantum
mechanically. $\hat{\mathcal{E}} (\hat{\mathcal{E}}^{\dagger})$ are the positive (negative) frequency
components of the quantized electric field
\begin{eqnarray}
\label{electric-field}
\hat{E}(\br,t)=\hat{\mathcal{E}}(\br,t)+\hat{\mathcal{E}}^\dag(\br,t) ~\mbox{,}
\end{eqnarray}
with
\begin{eqnarray}
\label{eq_fieldop}
\hat{\mathcal{E}}(\br,t)=\sum_s \left(\frac{2\pi\hbar\omega_s}{\Omega}\right)^{1/2} \hat{a}_s~
\mbox{e}^{i\bk_s\cdot\br-i\omega_s t} ~\mbox{,}
\end{eqnarray}
and $\Omega$ is the quantization volume.

In the rotating wave approximation (RWA) the molecule-field interaction is given by
\begin{equation}
\label{interaction}
\hat{H}_{int}(t)=\hat{\mathcal{E}}(\br,t)\hat{V}^{\dagger }+\hat{\mathcal{E}}^\dag(\br,t)\hat{V} ~\mbox{,}
\end{equation}
where we have partitioned the dipole operator $\hat{\mu}$ as $\hat{\mu} =\hat{V}+\hat{V}^{\dagger }$. Here $\hat{V}^{\dagger }(\hat{V})$ are the creation (annihilation) operators for excitations. The time-dependence in Eq. (\ref{interaction}) is in the interaction picture with respect to $\hat{H}_F$. Consequently, the time dependence of the total (field+molecule) density operator $\hat{\rho}_T(t)$ is governed by the Hamiltonian
\begin{equation}
\label{eq_ham}
\hat{H}(t)=\hat{H}_0+\hat{H}_{int}(t) ~\mbox{.}
\end{equation}

We assume that the field is initially in a coherent state,
\begin{eqnarray}
\label{coherent-state}
\vert \Psi_F \rangle= A_0 \mbox{exp}\left\{\sum_s \hat{a}_s^\dag\alpha_s\right\}|0\rangle ~\mbox{,} 
\end{eqnarray}
where
\begin{equation}
A_0=\mathrm{exp}(\sum_s \vert \alpha_s \vert^2) ~\mbox{,}
\end{equation}
is the normalization such that $\langle\Psi_F|\Psi_F\rangle=1$.
In Eq. (\ref{coherent-state}) $\alpha_s$ is the eigenvalue of $\hat{a}_s$, $\hat{a}_s|\Psi_F\rangle=\alpha_s|\Psi_F\rangle$, and
$|0\rangle$ is the vacuum state. 
The expectation value of the field is then
\begin{eqnarray} \label{field-trace}
\langle \Psi_F \vert \hat{E}(\br,t) \vert \Psi_F \rangle=\mathcal{E}(\br,t) + c.c. ~\mbox{,}
\end{eqnarray}
where
\begin{eqnarray}
\label{eps}
\mathcal{E}(\br,t)=\sum_s\left(\frac{2\pi\hbar\omega_s}{\Omega}\right)^{1/2}\mbox{e}^{i\bk_s\cdot\br-i\omega_st}\alpha_s ~\mbox{,}
\end{eqnarray}
is the field amplitude at space-point $\br$.

We shall define the time and frequency resolved signal $S(t)$ as the rate of change of the photon occupation, i.e.
\begin{eqnarray}
\label{eq_defsig}
S(t)& := &\frac{\mathrm{d}}{\mathrm{d}t} \expvdt{\hat{\mathcal{N}}} ~\mbox{,} 
\end{eqnarray}
where 
\begin{equation} \label{eq_defsig_1}
\hat{\mathcal{N}}:=\sum_{s} \hat{a}^{\dagger}_{s} \hat{a}_{s} ~\mbox{,}
\end{equation}
denotes the photon number operator. This definition applies to both spontaneous and stimulated process, as will be demonstrated later. In Eq. (\ref{eq_defsig_1}) the sum extends over all modes of interest in the experiment under consideration.

We shall adopt of the following convention for the various ensemble averages, one of which has been used in Eq. (\ref{eq_defsig}). In Eq. (\ref{eq_defsig}), $\expvdt{\dots}$ stands for a trace with respect to the density operator of the molecule where $\hat H_{int}$ includes {\em{all}} field modes (``total'' system). According to the definition of the signal, Eqs. (\ref{eq_defsig}) and (\ref{eq_defsig_1}), the field modes can be partitioned into detected (or signal) modes and the incoming modes. For later use, we introduce the average $\expvd{\dots}$, which denotes a trace with respect to the density operator of the molecule calculated in the interaction picture where $\hat H_{int}$ only includes the incoming modes. Averages with respect to a noninteracting system, where $\hat{H}_{int}$=0, will be denoted by $\expv{\dots}$. This corresponds to the limit $t \rightarrow -\infty$ (adiabatic switching).

The expectation value in Eq. (\ref{eq_defsig}) may be conveniently evaluated by switching to the Heisenberg picture, using the basic identity
\begin{eqnarray}
\label{eq_heisenberg}
\frac{\mathrm{d}}{\mathrm{d}t} \expvdt{\hat{\mathcal{N}}} & \equiv & \expvt{\frac{\mathrm{d}}{\mathrm{d}t} \hat{\mathcal{N}}_H} \nonumber \\
 & = & \expvt{ \sum_s\frac{i}{\hbar }[\hat{H}_{int}(t),\hat{a}_{s,H}^{\dagger }\hat{a}_{s,H}]} ~\mbox{,}
\end{eqnarray}
and the canonical commutation relations for $\hat{a}_s$ ($\hat{a}_s^{\dagger}$).  In Eq. (\ref{eq_heisenberg}) we denote operators in the Heisenberg picture by a subscript ``H''.

Evaluating the commutator in Eq. (\ref{eq_heisenberg}) and transforming back to the original interaction picure, we finally obtain for the signal, Eq. (\ref{eq_defsig}),
\begin{equation}
\label{signal-2}
S(t)=-\frac{2}{\hbar }\mbox{Im} \left \lbrace \expvdt{\hat{\mathcal{E}}(\br,t)\hat{V}^{\dagger }} \right \rbrace ~\mbox{.}
\end{equation}

Taking into account the time dependence of the total (system + field) density operator, which is determined by the Hamiltonian of Eq. (\ref{eq_ham}), the trace on the rhs of Eq. (\ref{signal-2}) can be computed
perturbatively in the field $E(\br,t)$ by switching to the interaction picture with respect to the Hamiltonian $\hat{H}_0$. We shall compute this trace using superoperators in Liouville space. To this end we associate with each operator $\hat{A}$ in Hilbert space, a left (L) and a right (R) operation in Liouville space, by defining
\begin{eqnarray}
\label{eq_defLR}
\hat{A}_L \hat{X} & := & \hat{A} \hat{X} ~\mbox{,} \nonumber \\
\hat{A}_R \hat{X} & := & \hat{X} \hat{A} ~\mbox{,} 
\end{eqnarray}
where $\hat{X}$ denotes an arbitrary Hilbert space operator. Furthermore, we introduce the following linear combinations of $L/R$ operations, which will be referred to as $+/-$ operations
\begin{eqnarray}
\label{definition+-}
\hat{A}_{\pm} & := & \frac{1}{\sqrt{2}}[\hat{A}_L \pm \hat{A}_R] ~\mbox{.}
\end{eqnarray}

A key tool in the following manipulations is the time-ordering operator in Liouville
space, $\mathcal{T}$; when acting on a product of superoperators, it
reorders them such that their time arguments increase from right to
left. In the interaction picture Eq. (\ref{signal-2}) can be expressed
as
\begin{eqnarray}
\label{signal-2a} S(t)&=&- \frac{2}{\hbar}
\mbox{Im}\left[\expvt{\mathcal{T} \hat{\mathcal{E}}_{L}(\br,t)\hat{V}_{L}^{\dagger}(t) \mbox{exp}\left\{-\frac{i}{\hbar}\int_{-\infty}^t \mathrm{d} \tau \sqrt{2}{\cal H}_{int-}(\tau)\right\}}\right] ~\mbox{,}
\end{eqnarray}
where ${\cal H}_{int-}$ is the superoperator corresponding to
$\hat{H}_{int}$, Eq. (\ref{interaction}). 

Note that $\hat{H}_{int}$ contains two interaction terms, one with the incoming field modes $\hat{H}^\prime_{int}$ and the other with the signal modes $\hat{H}''_{int}$. As before, the explicit time dependence represents the interaction picture
\begin{eqnarray}
\label{time-interaction}
\hat{a}_{sX}(t)&:=&\mbox{e}^{\frac{i}{\hbar}{\cal H}_{FX}t} \hat{a}_{sX} \mbox{e}^{-\frac{i}{\hbar}{\cal H}_{FX}t} ~\mbox{,} \nonumber\\
\hat{V}_X(t)&:=&\mbox{e}^{\frac{i}{\hbar}{\cal H}_{0X}t} \hat{V}_{X} \mbox{e}^{-\frac{i}{\hbar}{\cal H}_{0X}t} ~\mbox{,}
\end{eqnarray}
where $X=L,R$ and ${\cal H}_{FX}$ (${\cal H}_{0X}$) is the
superoperator corresponding to $\hat{H_F}(\hat{H}_0)$.

To lowest order in $\hat{H}''_{int}$ Eq. (\ref{signal-2a}) gives
\begin{eqnarray}
\label{signal-3}
S(t)&=& \frac{2\sqrt{2}}{\hbar^2}\mbox{Re}\sum_s \left(\frac{2\pi\hbar\omega_s}{\Omega}\right)^{1/2}
\mbox{e}^{i\bk_s\cdot\br}\nonumber\\
&\times&\int_{-\infty}^t \mathrm{d} \tau \expvd{ \mathcal{T} \hat{a}_{sL}(t)\hat{V}^\dag_L(t)\mathcal{H}''_{int-}(\tau)} ~\mbox{.}
\end{eqnarray}

The superoperator expression corresponding to Eq. (\ref{interaction}) gives
\begin{eqnarray}
\label{hint-lr}
\sqrt{2}\mathcal{H}_{int-}(\tau)&=& \hat{\mathcal{E}}_L(\br,\tau)\hat{V}^\dag_L(\tau) + \hat{\mathcal{E}}_L^\dag(\br,\tau)\hat{V}_L(\tau)\nonumber\\
&-&V^\dag_R(\tau)\mathcal{E}_R(\br,\tau)-V_R(\tau)\mathcal{E}_R^\dag(\br,\tau) ~\mbox{.}
\end{eqnarray}
Substituting in Eq. (\ref{signal-3}) and factorizing the correlation functions into products of
field and system parts, we get
\begin{eqnarray}
\label{signal-4}
& S(t)= \frac{4\pi\hbar}{\Omega}\mbox{Re}\sum_{s\sp} (\omega_s\omega_\sp)^{1/2}
\mbox{e}^{i(\bk_s-\bk_\sp)\cdot\br}\nonumber\\
\times & \int_{-\infty}^t d\tau \left[d_{LR}^{s\sp}(t,\tau)D_{RL}(\tau,t)
-d_{LL}^{s\sp}(t,\tau)D_{LL}(\tau,t)\right] ~\mbox{,} \nonumber\\
& & 
\end{eqnarray}
where we have introduced the field and system nonequilibrium Green's functions, $d_{XY}$ and $D_{XY}$, 
\begin{eqnarray}
\label{green}
d_{XY}^{s\sp}(t,\tau) & := & -\frac{i}{\hbar} \expvd{ \mathcal{T} \hat{a}_{sX}(t) \hat{a}_{\sp Y}^{\dagger}(\tau)} ~\mbox{,} \nonumber\\
D_{XY}(t,\tau)& := & -\frac{i}{\hbar} \expvd{\mathcal{T} \hat{V}_{X}(t) \hat{V}_{Y}^\dag(\tau)} ~\mbox{.}
\end{eqnarray}
In Eq. (\ref{signal-4}) we assume that the system is in its ground state for $t \rightarrow -\infty$, hence the following system correlation functions vanish
\begin{eqnarray}
\expvd{\mathcal{T} \hat{V}_L^{\dagger}(t) \hat{V}_L^{\dagger}(\tau)}  =  0 ~ \mbox{,} & \nonumber  \\
\expvd{\mathcal{T} \hat{V}_L^{\dagger}(t) \hat{V}_R^{\dagger}(\tau)} = 0 ~\mbox{,} &  ~t > \tau ~\mbox{.}
\end{eqnarray}

From the basic definitions of the Green's functions we have
\begin{eqnarray}
\label{green-prop}
D_{LL}(\tau,t) & = & \theta(\tau-t)D_{RL}(\tau,t)+\theta(t-\tau) D_{LR}(\tau,t) ~\mbox{,} \nonumber\\
d_{LL}(t,\tau) & = & \theta(t-\tau)d_{RL}(t,\tau)+\theta(\tau-t) d_{LR}(t,\tau) ~\mbox{.} \nonumber \\ & & 
\end{eqnarray}
Substituting Eq. (\ref{green-prop}) in (\ref{signal-4}) gives
\begin{eqnarray}
\label{signal-5}
& S(t) = \frac{4\pi\hbar}{\Omega}\mbox{Re}\sum_{s\sp} (\omega_s\omega_\sp)^{1/2}
\mbox{e}^{i(\bk_s-\bk_\sp)\cdot\br} & \nonumber\\
\times & \int_{-\infty}^t \mathrm{d}\tau \left[d_{LR}^{s\sp}(t,\tau)D_{RL}(\tau,t)
-d_{RL}^{s\sp}(t,\tau)D_{LR}(\tau,t)\right]  \mbox{.} & \nonumber \\ & &
\end{eqnarray}
Using Eqs. (\ref{coherent-state}) and (\ref{green}), the Green's functions for the field are given by
\begin{eqnarray}
\label{green-field}
d_{LR}^{s\sp}(t,\tau)&=& -\frac{i}{\hbar} \expvd{\hat{a}_\sp^\dag(\tau) \hat{a}_s(t)} \nonumber\\
&=&-\frac{i}{\hbar} \mbox{e}^{-i\omega_s t}\mbox{e}^{i\omega_\sp \tau}\alpha_s\alpha_\sp^* ~\mbox{,} \\
d_{RL}^{s\sp}(t,\tau)&=& -\frac{i}{\hbar}\expvd{ \hat{a}_s(t) \hat{a}_\sp^\dag(\tau)} \nonumber\\
&=& -\frac{i}{\hbar}\mbox{e}^{-i\omega_s t}\mbox{e}^{i\omega_\sp \tau}(\delta_{s\sp}+\alpha_s\alpha_\sp^*) ~\mbox{.}
\end{eqnarray}

The signal, Eq. (\ref{signal-5}), then becomes
\begin{eqnarray}
\label{signal-6}
S(t)&=& \frac{4\pi\hbar}{\Omega}\mbox{Im}\sum_{s\sp} (\omega_s\omega_\sp)^{1/2}
\mbox{e}^{i(\bk_s-\bk_\sp)\cdot\br}
\mbox{e}^{-i\omega_st}\nonumber\\
& \times &\int_{-\infty}^t d\tau \mbox{e}^{i\omega_\sp\tau}\left[\frac{}{}\alpha_s\alpha_\sp^*
D_{RL}(\tau,t) - (\delta_{s\sp}+\alpha_s\alpha_\sp^*)D_{LR}(\tau,t)\frac{}{}\right] ~\mbox{.}
\end{eqnarray}
In Eq. (\ref{signal-6}), the term with $\delta_{ss^{\prime}}$ represents spontaneous emission, whereas the terms proportional to $\alpha_s\alpha_{s^{\prime}}^{\ast}$ correspond to stimulated processes. Note also that the Green's functions $D_{XY}$ contain the density matrix of the driven system which involves all interactions with the incoming field.  

Equation (\ref{signal-6}) can also be written as
\begin{eqnarray}
\label{signal-7a}
& S(t)= 2 \mbox{Im} \int_{-\infty}^t \mathrm{d} \tau
\left[D_{RL}(\tau,t)
-D_{LR}(\tau,t)\right] \mathcal{E}(\br,t) \mathcal{E}^*(\br,\tau) &
\nonumber\\
& -  \frac{4\pi\hbar}{\Omega}\sum_s\omega_s
\mbox{Im}\int_{-\infty}^t d\tau \mbox{e}^{-i\omega_s(t-\tau)}D_{LR}(\tau,t) \mbox{,} & \nonumber \\
\end{eqnarray}
or equivalently
\begin{eqnarray}
\label{signal-7}
& S(t)= 2 \mbox{Im}\int_{-\infty}^t d\tau
\left[D_{RL}(\tau,t) \mathcal{E}(\mathbf{r},t) \mathcal{E}^{*}(\mathbf{r},\tau) - D_{LR}(t,\tau) \mathcal{E}^{*}(\mathbf{r},t) \mathcal{E}(\mathbf{r},\tau) \right]  & \nonumber\\
& -  \frac{4\pi\hbar}{\Omega}\sum_s\omega_s
\mbox{Im}\int_{-\infty}^t d\tau \mbox{e}^{i\omega_s(t-\tau)} D_{LR}(t,\tau) ~\mbox{.} & \nonumber \\
\end{eqnarray}
In Eq. (\ref{signal-7}) we made use of the symmetry relation
\begin{equation}
D_{LR}^{*}(\tau,t) = - D_{LR}(t,\tau) ~\mbox{,}
\end{equation}
to arrive at an expression, where interactions at time $t$ always occur from the left (i.e. on the ket). This choice will be adopted in the diagrammatic representation, which will be introduced in Sec. \ref{sec_sle}.

Equations (\ref{signal-7a}) or equivalently (\ref{signal-7}) constitute our basic NEGF expression for incoherent optical signals. The first term in the brackets describes the creation of excitations in the system by absorbing photons from the field and the second term represents the reverse process. Both processes are stimulated. The signal is thus given by the net photon flux. The last term represents spontaneous emission. In the coming two sections we shall apply these expressions to compute different signals.

\section{Diagrammatic representation for Spontaneous Light Emission (SLE)} \label{sec_sle}

To describe spontaneous light emission (SLE), which is an incoherent process, we assume that the scattered field is initially in its vacuum state and is generated by spontaneous emission. Thus $\mathcal{E}(\br,t)=0$ and Eq. (\ref{signal-7}),
reduces to
\begin{eqnarray}
\label{spon-signal-1}
S_{\mathrm{SLE}}(t)=-\frac{4\pi\hbar}{\Omega} \omega_s \mbox{Im}
\int_{-\infty}^t d\tau \mbox{e}^{i\omega_s(t-\tau)}D_{LR}(t,\tau) ~\mbox{,}
\end{eqnarray}
where it is furthermore assumed that the scattered field has only one mode $\omega_s$ thus
dropping the sum over $s$ (signal modes). By integrating over modes we can replace $1/\Omega$ by the density of modes.

Expanding the Green's function $D_{LR}$ to second order in the incoming field yields
\begin{eqnarray} \label{eq_spont}
& S_{\mathrm{SLE}}(t) = - \dfrac{4 \pi}{\Omega\hbar^2}\omega_{s} \mathrm{Re} \int_{-\infty}^{t}\mathrm{d}\tau \int \mathrm{d}\tau_{1} \int\mathrm{d}\tau_{2} \mathrm{e}^{i \omega_s(t-\tau)} \nonumber \\
& \times \expv{\mathcal{T} \hat{V}_{L}(t) \hat{V}_{R}^{\dagger}(\tau) \mathcal{H}_{int-}(\tau_{1}) \mathcal{H}_{int-}(\tau_{2})} ~\mbox{.}
\end{eqnarray}
In order to go any further, we need to expand the four-point correlation function of Eq. (\ref{eq_spont}) in $L/R$ operations. A systematic, diagrammatic technique for selecting the terms in Eq. (\ref{eq_spont}) that contribute within the RWA, will be introduced next.

We start with Eq. (\ref{eq_spont}), where the correlation function originates from expanding the Green's function $D_{LR}(t,\tau)$ in Eq. (\ref{spon-signal-1}) to second order in the incoming field. Hence in what follows, we associate the time variables $\tau_{1}$ and $\tau_{2}$ with the incoming field, whereas the variables $t$ and $\tau$ correspond to the signal field. We also note that thanks to the time ordering operator, Eq. (\ref{eq_spont}) is symmetric with respect to interchanging the dummy variables $\tau_{1}$ and $\tau_{2}$.

In order to derive an explicit expression for this correlation function in terms of superoperators, we first note that the interactions at times $t$ and $\tau$ both represent photon emission, one on the ket at time $t$, the other on the bra at time $\tau$. Since the system is assumed to be initially in its ground state, both emission events have to be preceded by two absorptions on either side. Although this implies a certain order of interactions on both the ket and the bra individually, their relative ordering in physical time remains a priori undetermined. This {\em{partial time ordering}} can be expressed diagrammatically by arranging the interactions on a loop.
We adopt the following general rules for constructing and reading the diagrams:
\begin{enumerate}
\item Time runs along the loop clockwise from bottom left to bottom right. \label{rule_time}
\item The left strand of the loop represents the ket, the right corresponds to the bra.
\item Each interaction with a field mode is represented by a wavy line on either the right (R-operators) or the left (L-operators).
\item The field is indicated by dressing the wavy lines with arrows, where an arrow pointing to the right represents the field annihilation operator $\hat{\mathcal{E}}(\mathbf{r},t)$, which involves the term $\mathrm{e}^{i(\mathbf{k}_{s} \cdot \mathbf{r} - \omega_{s}t)}$ (see Eq. (\ref{eq_fieldop})). Conversely, an arrow pointing to the left corresponds to the field creation operator $\hat{\mathcal{E}}^{\dagger}(\mathbf{r},t)$, being associated with $\mathrm{e}^{-i(\mathbf{k}_{s} \cdot \mathbf{r} - \omega_{s}t)}$. This is made explicit by adding the wavevectors $\pm \mathbf{k}_{s}$ to the arrows.
\item Within the RWA (Eq. (\ref{interaction})), each interaction with $\hat{\mathcal{E}}(\mathbf{r},t)$ is accompanied by applying the operator $V^{\dagger}$, which leads to excitation of the state represented by ket and deexcitation of the state represented by the bra, respectively. Arrows pointing ``inwards'' (i.e. pointing to the right on the ket and to the left on the bra) consequently cause absorption of a photon by exciting the system, whereas arrows pointing ``outwards'' (i.e. pointing to the left on the bra and to the right on the ket) represent deexciting the system by photon emission.
\item The interaction at the observation time $t$, is fixed and is always the last. As a convention, it is chosen to occur from the left. This can always be achieved by a reflection of all interactions through the center line between the ket and the bra, which corresponds to taking the complex conjugate of the original correlation function.
\item Interactions within each strand are time-ordered, but interactions on different strands are not. Each loop can be further decomposed into several fully-time-ordered diagrams (double sided Feynman diagrams). These can be generated from the loop by simply shifting the arrows along each strand, thus changing their position relative to the interactions on the other strand. Each of these relative positions then gives rise to a particular fully time-ordered diagram. 
\item The overall sign  of the correlation function is given by $(-1)^{N_R}$, where $N_{R}$ stands for the number of interactions from the right, at times $\tau_{1}$ and $\tau_{2}$.
\end{enumerate}

\begin{figure}
\begin{center}
\includegraphics[width=0.6\columnwidth]{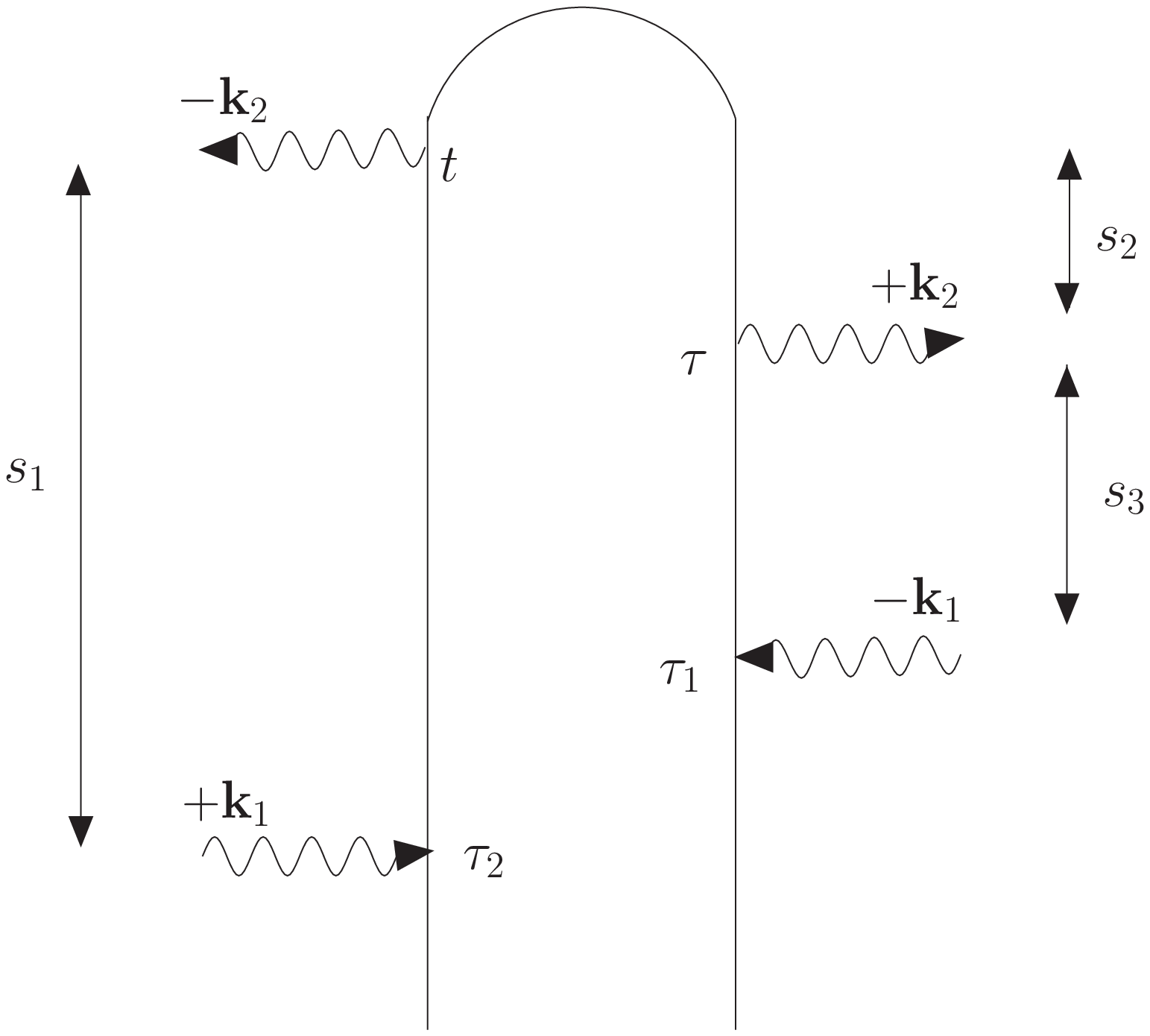}
\caption{Loop diagram for SLE. $\mathbf{k_1}$ is the incoming field, and $\mathbf{k}_2$ is the signal field. Note that the interactions are time ordered within each strand, but not between strands. $s_1, s_2, s_3$ are the delay times between the interactions along the loop.}
\label{fig_SLE_loop_1}
\end{center}
\end{figure}

We note that loop diagrams drawn according to the rules presented above, lead to double sided Feynman diagrams that follow the standard conventions employed within the semiclassical theory of nonlinear optics \cite{mukbook}. Using these rules, the SLE is represented by the single loop diagram displayed in Fig. \ref{fig_SLE_loop_1}. We denote the incoming field as $\pm \mathbf{k}_{1}$ and the signal field as $\pm \mathbf{k}_{2}$. The loop translates Eq. (\ref{eq_spont}) into the following expression for the signal,
\begin{eqnarray} \label{eq_spont_1}
S_{\mathrm{SLE}}(t) =  \dfrac{4\pi}{\Omega\hbar^2}\omega_{2} \mathrm{Re}\int_{-\infty}^{t}\mathrm{d}\tau \int \mathrm{d}\tau_{1} \int\mathrm{d}\tau_{2} \mathrm{e}^{i \omega_2 (t-\tau)}  \nonumber \\
\expv{\mathcal{T} \hat{V}_L(t) \hat{V}_R^{\dagger}(\tau) \hat{V}_L^{\dagger}(\tau_2) \hat{V}_R(\tau_1)} \expv{\mathcal{T}\hat{\mathcal{E}}_L(\mathbf{r},\tau_2) \hat{\mathcal{E}}_R^{\dagger}(\mathbf{r},\tau_1)}  ~\mbox{,}  \end{eqnarray}
where an additional prefactor of 2 was added to take into account the symmetry with respect to the dummy variables $\tau_{1}$ and $\tau_{2}$. 

\begin{figure}
\begin{center}
\includegraphics[width=\columnwidth]{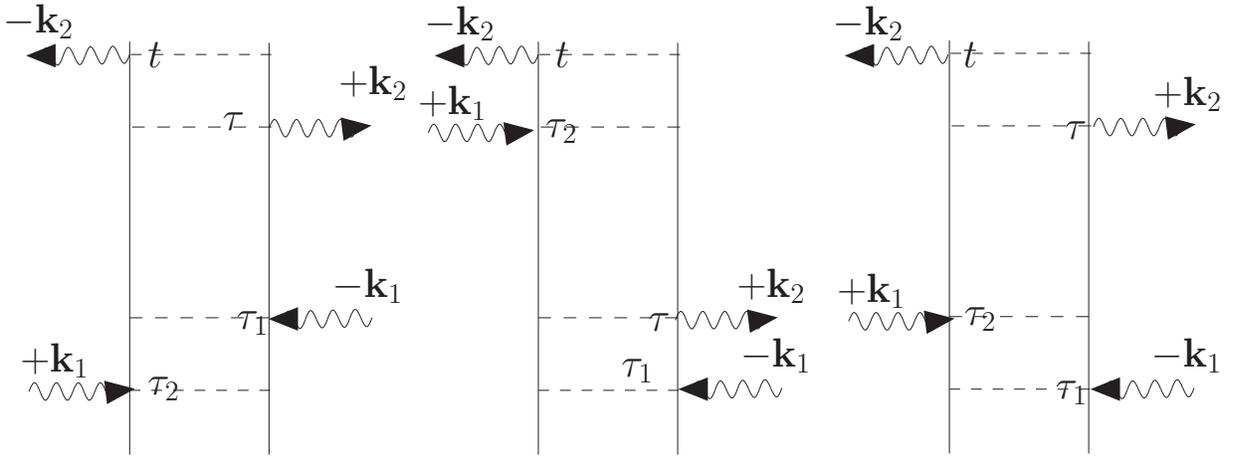}
\caption{Time-ordered Feynman diagrams of SLE, generated by shifting the arrows of Fig. \ref{fig_SLE_loop_1} along each strand thus changing their relative time ordering. Each of the possible three relative positions then gives one fully time ordered diagram (double sided Feynman diagram).}
\label{fig_SLE_loop_2}
\end{center}
\end{figure}

If desired, this can further be decomposed into fully time-ordered terms using the double sided diagrams of Fig. \ref{fig_SLE_loop_2}. As indicated above we can now e.g. shift the arrow in Fig. \ref{fig_SLE_loop_1} corresponding to time $\tau_{2}$ along the ket, thus obtaining 3 possible relative positions with respect to the interactions at times $\tau$ and $\tau_{1}$. The loop diagram consequently splits into 3 double sided Feynman diagrams. These are depicted in Fig. \ref{fig_SLE_loop_2}.

The resulting fully-time-ordered expression for the signal reads
\begin{eqnarray} \label{eq_spont_2}
& S_{\mathrm{SLE}}(t) =  \dfrac{4\pi}{\Omega\hbar^2}\omega_{2} \mathrm{Re}\int_{-\infty}^{t}\mathrm{d}\tau \int \mathrm{d}\tau_{1} \int\mathrm{d}\tau_{2} \mathcal{E}_1^{*}(\mathbf{r},\tau_1) \mathcal{E}_1(\mathbf{r},\tau_2) & \nonumber \\
& \mathrm{e}^{i \omega_s (t-\tau)} \left[ \theta(t \tau) \theta(\tau \tau_1) \theta(\tau_1 \tau_2) \expv{V_L(t) V_R^{\dagger}(\tau) V_R(\tau_1) V_L^{\dagger}(\tau_2)} \right. & \nonumber \\ 
& + \theta(t \tau_2) \theta(\tau_2 \tau) \theta(\tau \tau_1) \expv{\hat{V}_L(t) \hat{V}_L^{\dagger}(\tau_2) \hat{V}_R^{\dagger}(\tau) \hat{V}_R(\tau_1)} & \nonumber \\ 
& + \theta(t \tau) \theta(\tau \tau_2) \theta(\tau_2 \tau_1) \expv{V_L(t) V_R^{\dagger}(\tau) V_L^{\dagger}(\tau_2) V_R(\tau_1)} \left. \right] ~\mbox{.} &
\end{eqnarray}
Here, we have made use of Eq. (\ref{coherent-state}) to evaluate the correlation functions for the incoming field, since
\begin{eqnarray}
& \expv{\mathcal{T} \hat{\mathcal{E}}_L(\mathbf{r},\tau_2) \hat{\mathcal{E}}_R^{\dagger}(\mathbf{r},\tau_1)}=\dfrac{2\pi\hbar}{\Omega} \omega_1 \vert \alpha_1 \vert^2 \mathrm{e}^{- i\omega_1(\tau_1-\tau_2)} & \nonumber \\
& = \mathcal{E}_1^{*}(\mathbf{r},\tau_1) \mathcal{E}_1(\mathbf{r},\tau_2) ~\mbox{.}
\end{eqnarray}
Note that, in Eq. (\ref{eq_spont_2}) full time-ordering is expressed by adding a product of step functions to each of the contributing terms. We apply the short hand notation $\theta(\tau_1 \tau_2)=\theta(\tau_1-\tau_2)$.

Next, we calculate the signal for a frequency-domain experiment with stationary beams. This will also illustrate how to extend the diagrammatic rules to frequency-domain experiments. We adopt the following convention for the Fourier transform of a function $f$
\begin{eqnarray} \label{eq_deffour_fun}
\tilde{f}(\omega) & = & \int \mathrm{d}t \mathrm{e}^{i \omega t} f(t) ~\mbox{,} \nonumber \\
f(t) & = & \frac{1}{2 \pi}\int \mathrm{d} \omega \mathrm{e}^{-i \omega t} \tilde{f}(\omega) ~\mbox{.}
\end{eqnarray}

When the incoming field is stationary, the integrals in Eq. (\ref{eq_spont_1}) can be evaluated directly, which by application of Eq. (\ref{eq_deffour_fun}) leads to a multiple Fourier transform of the partially ordered 4-point system correlation function in Eq. (\ref{eq_spont_1}). To simplify this calculation, it is advantageous to switch to a new set of time variables (``$s$-variables''),  $s_1 :=  t - \tau_2$, $s_2  :=  t - \tau$, $s_3  :=  \tau - \tau_1$, which represent the time intervals of interactions {\em{along}} the loop. This choice becomes intuitive when looking at the diagram, Fig. \ref{fig_SLE_loop_1}.

We can now show that the correlation function in Eq. (\ref{eq_spont_2}) depends on the $s$-variables in a simple way. To this end, we transform the system correlation function in Eq. (\ref{eq_spont_1}) to a form that only involves $L$-superoperators, hence
\begin{eqnarray} \label{eq_slecorr}
& \expv{\mathcal{T} \hat{V}_L(t) \hat{V}_R^{\dagger}(\tau) \hat{V}_R(\tau_1) \hat{V}_L^{\dagger}(\tau_2)} & \nonumber \\
& =\expv{\hat{V}_L(\tau_1) \hat{V}_L^{\dagger}(\tau) \hat{V}_L(t) \hat{V}_L^{\dagger}(\tau_2)} & 
\end{eqnarray}
In terms of the $s$-variables, Eq. (\ref{eq_slecorr}) becomes
\begin{eqnarray} \label{eq_corrfun}
& \expv{\mathcal{T} \hat{V}_L(t) \hat{V}_R^{\dagger}(\tau) \hat{V}_R(\tau_1) \hat{V}_L^{\dagger}(\tau_2)} & \nonumber \\
& = \expv{\hat{V}_L \mathcal{G}^{\dagger}(s_3) \hat{V}_L^{\dagger} \mathcal{G}^{\dagger}(s_2) \hat{V}_L \mathcal{G}(s_1) \hat{V}_L^{\dagger}} ~\mbox{,} & 
\end{eqnarray}
where we define the Liouville space propagator
\begin{eqnarray}
\mathcal{G}(s) & := & \theta(s) \mathrm{exp}\left(-\frac{i}{\hbar} \sqrt{2} \mathcal{H}_{0-}s \right) ~\mbox{.}
\end{eqnarray}
Note that, when associating all interactions with the ket, one needs to propergate twice backwards in time. This amounts to applying the retarded propagator $\mathcal{G}^{\dagger}$ twice. 

Using Eq. (\ref{eq_corrfun}) the integration in Eq. (\ref{eq_spont_1}), leads to a multiple Fourier transform of the correlation function in terms of the loop (s)- variables. This is readily calculated, giving the following expression for the frequency-domain SLE signal
\begin{eqnarray} \label{eq_SLE_freq}
& S_{\mathrm{SLE}}(\omega_1,\omega_2)=\dfrac{4 \pi}{\hbar^2 \Omega} \vert \mathcal{E}_{1,0} \vert^2 \omega_2 \nonumber  \\
& \mathrm{Im} \expv{\hat{V}_L \mathcal{G}^{\dagger}(\omega_1) \hat{V}_L^{\dagger} \mathcal{G}^{\dagger}(\omega_1-\omega_2) \hat{V}_L \mathcal{G}(\omega_1) \hat{V}_L^{\dagger}} \mbox{.} \nonumber \\
\end{eqnarray}

Since all interactions in Eq. (\ref{eq_SLE_freq}) are ``left''-operations that act on the ket, it is possible to express the final result for the signal in Hilbert space, giving
\begin{eqnarray} \label{eq_SLE_freq_hilb}
& S_{\mathrm{SLE}}(\omega_1,\omega_2)=\dfrac{4 \pi}{\hbar^2 \Omega} \vert \mathcal{E}_{1,0} \vert^2 \omega_2 \nonumber  \\
& \mathrm{Im} \expv{\hat{V} \hat{G}^{\dagger}(\omega_g + \omega_1) \hat{V}^{\dagger} \hat{G}^{\dagger}(\omega_g + \omega_1 - \omega_2) \hat{V} \hat{G}(\omega_g + \omega_1) \hat{V}^{\dagger}} \mbox{.} \nonumber \\
\end{eqnarray}
Upon conversion of Eq. (\ref{eq_SLE_freq}), the frequency-domain Liouville space propagators,
\begin{equation} \label{eq_prop_Liov}
\mathcal{G}(\omega)=\dfrac{\hbar}{\hbar\omega - \sqrt{2} \mathcal{H}_{0-} + i \eta} ~\mbox{,}
\end{equation}
are replaced by Hilbert-space propagators, 
\begin{equation} \label{def_prop_Hilb}
\hat{G}(\omega + \omega_g)=\dfrac{\hbar}{\hbar \omega + \hbar \omega_g - \hat{H}_{0} + i \eta} ~\mbox{.}
\end{equation}
Here, $\omega_g$ is the material frequency of the ground state, which accounts for the free evolution of the bra. In both Eqs. (\ref{eq_prop_Liov}) and (\ref{def_prop_Hilb}) the infinitesimal $\eta > 0$ arises from causality and guarantees the convergence of the Fourier transform.

Equation (\ref{eq_SLE_freq_hilb}) can be recast in the form
\begin{eqnarray}
& S_{\mathrm{SLE}}(\omega_1,\omega_2)=-\dfrac{2 \pi i}{\hbar^2 \Omega} \vert \mathcal{E}_{1,0} \vert^2 \omega_2 \nonumber \\
& \expv{\hat{V} \hat{G}^{\dagger}(\omega_g + \omega_1) \hat{V}^{\dagger} \left[ \hat{G}^{\dagger}(\omega_g + \omega_1-\omega_2) -\hat{G}(\omega_g + \omega_1-\omega_2) \right] \hat{V} \hat{G}(\omega_g + \omega_1) \hat{V}^{\dagger}} ~\mbox{,} \nonumber \\
\end{eqnarray}
which by using the level scheme of Fig. \ref{fig_3ls} (a) yields the Kramers-Heisenberg formula
\begin{eqnarray} \label{eq_kramers_1}
S_{\mathrm{SLE}}(\omega_1,\omega_2)=-\dfrac{4 \pi^2}{\hbar^2 \Omega} \omega_2 \vert \mathcal{E}_{1,0} \vert^2 \vert \chi_{ca} \vert^2 \dfrac{\Gamma_{ca} / \pi}{(\omega_1-\omega_2-\omega_{ca})^2 + \Gamma_{ca}^2} ~\mbox{,} \nonumber \\
\end{eqnarray}
where 
\begin{equation} \label{eq_kramers_2}
\chi_{ca}=\dfrac{\mu_{ab}\mu_{bc}}{\omega_1-\omega_{ba}+i\Gamma_{ba}} ~\mbox{.}
\end{equation}
In Eqs. (\ref{eq_kramers_1}) and (\ref{eq_kramers_2}) $\omega_{ca}=\omega_{c}-\omega_{a}$ denotes the transition frequency between the levels c and a and we have added a phenomenological dephasing rate $\Gamma_{ca}$.

By proceeding along the same line for an arbitrary loop diagram, one can establish the following rules, that allow to translate a given diagram into its frequency-domain expression. These complement the rules given earlier for time-domain expressions.
\begin{enumerate}
\item In the frequency-domain the loop translates into an alternating product of interactions (arrows) and periods of free evolutions (vertical solid lines) along the loop.
\item Since the loop time goes clockwise along the loop, periods of free evolution on the left amount to propagating forward in real time ($\hat{G}$), whereas evolution on the right corresponds to backward propergation ($\hat{G}^{\dagger}$). 
\item Each $\hat{G}$ adds the multiplicative factor $i$, whereas each $\hat{G}^{\dagger}$ results in a multiplication by $(-i)$.
\item Frequency arguments of each propagator are cumulative, i.e. they are given by the sum of all ``earlier'' interactions along the loop. The ground state frequency $\omega_g$ must be added to all Green's functions.
\end{enumerate}
Equation (\ref{eq_SLE_freq_hilb}) can be immediately generated from Fig. \ref{fig_PP_loop} by applying these rules.

\section{The Pump-Probe Signal} \label{sec_pumpprobe}

The simplest nonlinear optical technique involves two fields, $\mathbf{k}_1$ (the pump) and $\mathbf{k}_2$ (the probe). The signal is defined as the difference of the probe transmitted intensity with and without the pump.
We assume that the probe intensity is high so that spontaneous emission is negligible compared to stimulated emission. The second term in Eqs. (\ref{signal-7a}) and (\ref{signal-7}), which describes spontaneous emission, is thus neglected. Using Eq. (\ref{signal-7a}) and expanding to second order in the incoming field, the signal then reduces to
\begin{eqnarray}
\label{pp-signal-1a}
&& S_{\mathrm{PP}}(t)=2\frac{1}{\hbar^3}\mbox{Re} \int_{-\infty}^t d\tau \int d\tau_1\int d\tau_2\mathcal{E}_2(\br,t)\mathcal{E}_2^*(\br,\tau)\nonumber\\
&&\left[\expv{ \mathcal{T} \hat{V}_R(\tau) \hat{V}_L^\dag(t) {\cal H}_{int-}(\tau_1) {\cal H}_{int-}(\tau_2)} \right.\nonumber\\
&-&\left. \expv{\mathcal{T} \hat{V}_L(\tau) \hat{V}_R^\dag(t) {\cal H}_{int-}(\tau_1) {\cal H}_{int-}(\tau_2)} \right] ~\mbox{.}
\end{eqnarray}
Using the identity
\begin{equation}
D_{RL}(\tau,t)-D_{LR}(\tau,t)=\frac{i}{\hbar}\expvd{\mathcal{T} \hat{V}_{+}^{\dagger}(t) \hat{V}_{-}(\tau)} ~\mbox{,}
\end{equation}
we can express the pump-probe signal in terms of $+/-$ operators as
\begin{eqnarray}
\label{pp-signal-2}
S_{\mathrm{PP}}(t)&=& - \frac{2}{\hbar^3} \mbox{Re}\int_{-\infty}^t d\tau \int \mathrm{d}\tau_1 \int \mathrm{d}\tau_2
\mathcal{E}_2(\br,t)\mathcal{E}_2^*(\br,\tau)\nonumber\\
&&\expv{\mathcal{T} V_+^{\dagger}(t)V_-(\tau){\cal H}_{int-}(\tau_1){\cal H}_{int-}(\tau_2)} ~\mbox{.}
\end{eqnarray}
This is equivalent to the classical expression for the signal in terms of $\chi^{(3)}$. This becomes even more apparent if Eq. (\ref{pp-signal-2}) is rewritten in the form
\begin{eqnarray} \label{eq_pp_chi3}
S_{\mathrm{PP}}(t) = -\frac{2}{\hbar^3} \mathrm{Re} \int_{-\infty}^{t} \mathrm{d} \tau \int \mathrm{d}\tau_1 \int \mathrm{d}\tau_2 \mathcal{E}_2(\mathbf{r},t) \mathcal{E}_2^{*}(\mathbf{r},\tau) \nonumber \\ 
\mathcal{E}_1(\mathbf{r},\tau_1) \mathcal{E}_1^{*}(\mathbf{r},\tau_2) \expv{\mathcal{T} \hat{V}_{+}^{\dagger}(t) \hat{V}_{-}(\tau) \hat{V}_{-}^{\dagger}(\tau_1) \hat{V}_{-}(\tau_2)} ~\mbox{,}
\end{eqnarray}
which is equivalent to Eq. (\ref{pp-signal-2}), if only terms proportional to the field intensities are taken into account. The details are given in appendix \ref{app_chi3}.

The forgoing example of SLE illustrated the use of Keldysh-Schwinger loops to compute the signal. In particular, by using this {\em{partially time ordered}} diagrammatic representation, we were able to reduce the number of diagrams from three to one. The merits of this compact notation will become even more apparent for the pump-probe signal, as will be shown next. 

We start with an expression for the signal using Eq. (\ref{signal-7})
\begin{eqnarray}
\label{pp-signal-1}
& S_{\mathrm{PP}}(t)= 2\frac{1}{\hbar^3}\mbox{Re} \int_{-\infty}^t d\tau \int d\tau_1\int d\tau_2\nonumber\\
& \left[\mathcal{E}_2(\br,t) \mathcal{E}_2^*(\br,\tau) \expv{\mathcal{T} \hat{V}_R(\tau) \hat{V}_L^\dag(t) {\cal H}_{int-}(\tau_1) {\cal H}_{int-}(\tau_2)} \right.\nonumber\\
-&\left. \mathcal{E}_2^*(\br,t)\mathcal{E}_2(\br,\tau) \expv{\mathcal{T} \hat{V}_L(t) \hat{V}_R^\dag(\tau) {\cal H}_{int-}(\tau_1) {\cal H}_{int-}(\tau_2)} \right] ~\mbox{,}
\end{eqnarray}
where in analogy to Eq. (\ref{pp-signal-1a}), we neglect the spontaneous emission term.

\begin{figure}
\begin{center}
\subfigure[]{\includegraphics[width=0.3\columnwidth]{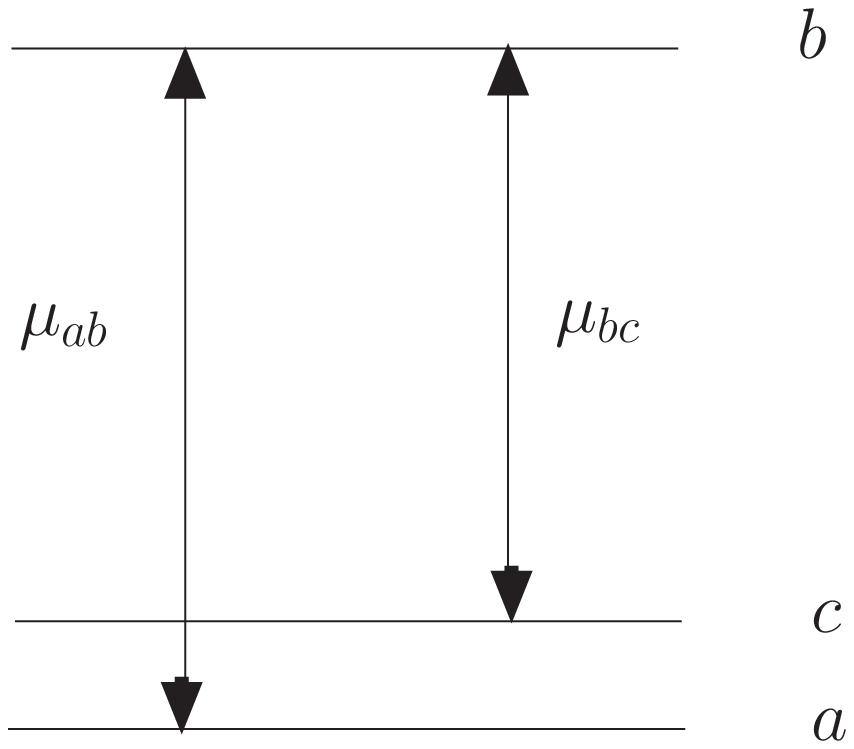}}
\subfigure[]{\includegraphics[width=0.3\columnwidth]{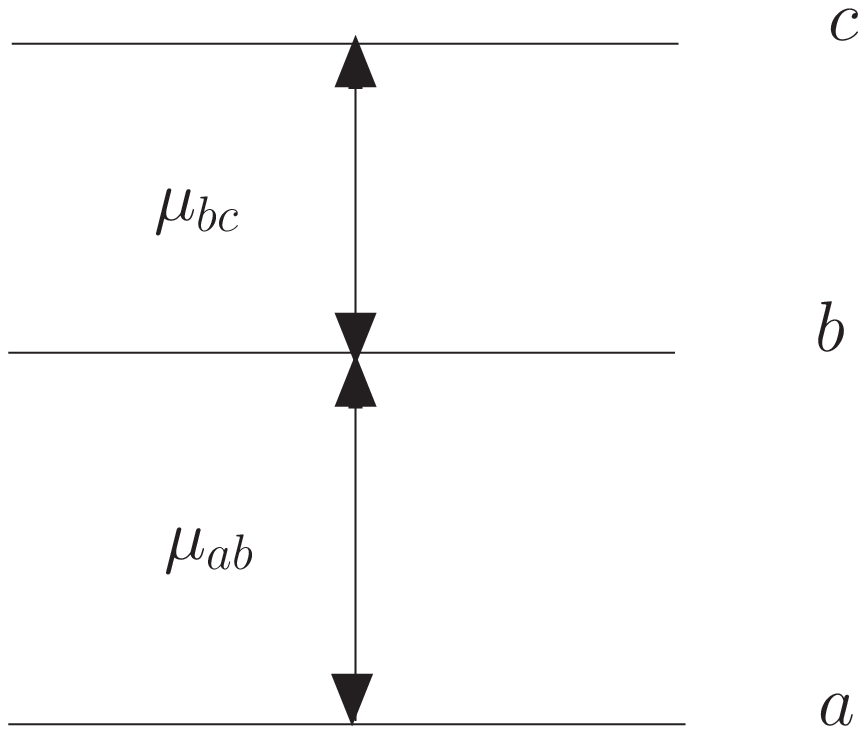}}
\caption{The three level systems with sequential dipole couplings used in the derivation of the Kramers Heisenberg relation for the SLE (panel (a); Eq. (\ref{eq_kramers_2})) and the pump-probe signal (panel (b)).}
\label{fig_3ls}
\end{center}
\end{figure}

We note that the second term in Eq. (\ref{pp-signal-1}) coincides with the SLE correlation function, Eq. (\ref{eq_spont}). What remains therefore is to resolve the first term in terms of $L/R$ operations. For simplicity we assume a three level ladder with sequential transition dipole moments as shown in Fig. \ref{fig_3ls} (b).

We again start our analysis by looking at the two interactions at times $t$ and $\tau$. Contrary to the second term in Eq. (\ref{pp-signal-1}), both of these correspond to absorptions rather than emissions. For the dipole selection rules shown in Fig. \ref{fig_3ls} (b), this leaves two possibilities for placing the remaining interactions along the loop: We either have three interactions on one side, or two interactions on either side (see Fig. \ref{fig_PP_loop}).

For the former case and the model of Fig. \ref{fig_3ls} (b), these can only correspond to two absorbtions and one emission. By applying the same argument used for SLE, the earliest of these three interactions must necessarily be an absorption. To proceed further, one then needs to distinguish whether these interactions occur on the ket or the bra. Since the last interaction is fixed to time $t$, the former case correspondingly allows only one possible loop diagram, which is displayed in panel (b) of Fig. \ref{fig_PP_loop}.

When placing the three integrations on the right strand, one has to allow for all possible orderings of their associated times $\tau$, $\tau_1$ and $\tau_2$. At this point we reiterate that when drawing a loop diagram, even though the relative time ordering of the ket and the bra interactions is unspecified, the temporal order within each strand is fixed. Hence having three interactions with the bra, generates 4 loop diagrams, which arise upon performing permutions of the arrows on the right side. Panels (e) - (h) of Fig. \ref{fig_PP_loop} show these diagrams. Note that in this case the loop diagrams are actually fully time ordered. 

We now turn to the diagrams with two left and two right interactions. Since the last interaction occurs on the ket at time $t$, for a system starting off in the ground state, the other interaction on the ket also must be an absorption. Finally, we note that when calculating the correlation functions Eq. (\ref{pp-signal-1}), one needs to take a trace in the end, which restricts both interactions on the bra to be absorptions as well. Arguing along this line leads to the two loop diagrams shown in panels (c) and (d) of Fig. \ref{fig_PP_loop}. The two loops again take into account the two possible permutations of placing interactions on the bra.

In summary, the pump-probe signal can be represented by eight loop diagrams, seven stem from the first correlation function in Eq. (\ref{pp-signal-1}), whereas the remaining one is SLE-type. The corresponding expression for the signal  derived using the time-domain rules is given in appendix \ref{app_pp}.

For the sake of completness, we shall compare the QED signal with the semiclassical result, where the pump-probe signal is calculated as a third order response in the direction $+ \mathbf{k}_1 - \mathbf{k}_1 + \mathbf{k}_2$. in  Note that the diagrams in panels (b) - (h) in Fig. \ref{fig_PP_loop}, correspond to $+ \mathbf{k}_1 - \mathbf{k}_1 - \mathbf{k}_2$. This difference can however easily be resolved by taking the complex conjugate of these diagrams, which leaves the expression for the signal in Eq. (\ref{eq_pp-signal-2a}) invariant. Pictorially, since this amounts to reflecting the arrows through the center line between the ket and bra, we recover the classical combination of wavevectors $+ \mathbf{k}_1 - \mathbf{k}_1 + \mathbf{k}_2$. As was the case for SLE, we can generate the corresponding fully-time ordered diagrams from the loops displayed in Fig. \ref{fig_PP_loop}. The resulting 16 double-sided Feynman diagrams are summarized in Fig. \ref{fig_PP_loop_open}. In addition, as has been shown in Sec. \ref{sec_sle} for SLE, frequency-domain expressions follow naturally from the loop diagrams, since the field permutations are already built in. Applying the rules given in Sec. \ref{sec_sle} to the diagrams for the pump-probe (Fig. \ref{fig_PP_loop}), we can immediately write down the frequency-domain signal 
\begin{widetext}
\begin{eqnarray} \label{eq_pp_frequ}
& S_{\mathrm{PP}}(\omega_1,\omega_2)=-\frac{4}{\hbar^3} \vert \mathcal{E}_{1,0} \vert^2 \vert \mathcal{E}_{2,0} \vert^2 \mathrm{Im} \left \lbrace  - \expv{\hat{V} \hat{G}^{\dagger}(\omega_g +\omega_1) \hat{V}^{\dagger} \hat{G}^{\dagger}(\omega_g +\omega_1-\omega_2) \hat{V} \hat{G}(\omega_g +\omega_1) \hat{V}^{\dagger}} \right. & \nonumber \\
& +  \expv{\hat{V} \hat{G}^{\dagger}(\omega_g + \omega_2) \hat{V}^{\dagger} \hat{G}(\omega_g +\omega_1-\omega_1) \hat{V} \hat{G}(\omega_g +\omega_1) \hat{V}^{\dagger}} \nonumber \\
& +  \expv{\hat{V} \hat{G}^{\dagger}(\omega_g +\omega_1) \hat{V} \hat{G}^{\dagger}(\omega_g +\omega_1+\omega_2) \hat{V}^{\dagger} \hat{G}(\omega_g +\omega_1) \hat{V}^{\dagger}} \nonumber \\
& +  \expv{\hat{V} \hat{G}^{\dagger}(\omega_g +\omega_2) \hat{V}  \hat{G}^{\dagger}(\omega_g +\omega_1+\omega_2) \hat{V}^{\dagger} \hat{G}(\omega_g +\omega_1) \hat{V}^{\dagger}} \nonumber \\
& +  \expv{\hat{V} \hat{G}^{\dagger}(\omega_g +\omega_1) \hat{V}^{\dagger} \hat{G}^{\dagger}(\omega_g +\omega_2-\omega_2) \hat{V} \hat{G}^{\dagger}(\omega_g +\omega_2) \hat{V}^{\dagger}} \nonumber \\
& +  \expv{\hat{V} \hat{G}^{\dagger}(\omega_g +\omega_2) \hat{V}^{\dagger} \hat{G}^{\dagger}(\omega_g +\omega_2-\omega_1) \hat{V} \hat{G}^{\dagger}(\omega_g +\omega_2) \hat{V}^{\dagger}}  \nonumber \\
& +  \expv{\hat{V} \hat{G}^{\dagger}(\omega_g +\omega_1) \hat{V} \hat{G}^{\dagger}(\omega_g +\omega_1+\omega_2) \hat{V}^{\dagger} \hat{G}(\omega_g +\omega_2) \hat{V}^{\dagger}} \nonumber \\
& +  \left. \expv{\hat{V} \hat{G}^{\dagger}(\omega_g +\omega_2) \hat{V} \hat{G}^{\dagger}(\omega_g +\omega_1+\omega_2) \hat{V}^{\dagger} \hat{G}^{\dagger}(\omega_g +\omega_2) \hat{V}^{\dagger}} \right \rbrace \mbox{.} \nonumber  \\
\end{eqnarray}
\end{widetext}
The eight terms correspond respectively to the eight diagrams in Fig. \ref{fig_PP_loop}.

\section{Coherent vs. incoherent nonlinear optical processes} \label{sec_cohincoh}

So far we have focused on incoherent processes. We could thus consider a single molecule and simply multiply the signal by $N$ in the end. To describe coherent signals we consider $\hat{V}$ in Eq. (\ref{signal-2}) as the dipole moment of a collection of noninteracting molecules at positions $\mathbf{r}_{\alpha}$, i.e.
\begin{equation} \label{eq_macrdipole}
\hat{V}=\sum_{\alpha} \hat{V}_{\alpha} \delta(\mathbf{r}-\mathbf{r}_{\alpha}) ~\mbox{,}
\end{equation}
where $\hat{V}_{\alpha}$ now denotes the dipole operator of a single molecule. Subsequent interactions of the detecting field can take place with the same ($\alpha = \beta$) or with different ($\alpha \neq \beta$) molecules. We shall describe an $(n+1)$-wave mixing experiment, where $n$ incoming modes generate a signal in a new direction. When the signal mode (index $s$) is initially in its vacuum state, the spontaneous emission signal can be  described by Eq. (\ref{signal-7a}) modified to include Eq. (\ref{eq_macrdipole}).
We then get
\begin{eqnarray} \label{eq_cohnwm_1}
& S_{\mathrm{T}} = \frac{4 \pi}{\Omega} \omega_s \mathrm{Re} \left \lbrace \int_{-\infty}^{+\infty} \mathrm{d} t \int_{-\infty}^{t} \mathrm{d} \tau \right. \nonumber \\
& \left. \sum_{\alpha,\beta} \mathrm{e}^{-i\omega_s(t-\tau)} \mathrm{e}^{i \mathbf{k}_s \cdot (\mathbf{r}_{\beta}-\mathbf{r}_{\alpha})} \expvd{\mathcal{T} \hat{V}_{\alpha,L}(\tau) \hat{V}_{\beta,R}^{\dagger}(t)} \right \rbrace ~\mbox{,} 
\end{eqnarray}
where we have further added a time-integration to describe a frequency-domain experiment \cite{Muk278}. Equation (\ref{eq_cohnwm_1}) is a general expression for spontaneously generated signals. The $\alpha = \beta$ and $\alpha \neq \beta$ terms in Eq. (\ref{eq_cohnwm_1}) give rise to incoherent ($S_I$) and coherent ($S_C$) signals, respectively. We thus write
\begin{equation} \label{eq_defcohincoh}
S_{\mathrm{T}}(\omega_s) = N \cdot S_I(\omega_s) + N(N-1) \cdot S_C(\omega_s) ~\mbox{,}
\end{equation}
where the incoherent signal is given by
\begin{eqnarray} \label{eq_sigincoh}
S_I(\omega_s) = \frac{4 \pi}{\Omega} \omega_s \mathrm{Re} \left \lbrace \int_{-\infty}^{+\infty} \mathrm{d} t \int_{-\infty}^{t} \mathrm{d} \tau \mathrm{e}^{-i\omega_s(t-\tau)} \expvd{\mathcal{T} \hat{V}_{\alpha,L}(\tau) \hat{V}_{\alpha,R}^{\dagger}(t)} \right \rbrace ~\mbox{.} \nonumber \\ 
\end{eqnarray}

For uncorrelated particles, we have $[\hat{V}_{\alpha},\hat{V}_{\beta}]=0$. The $\alpha \neq \beta$ sum may thus be factorized into $\alpha$ and $\beta$,
\begin{eqnarray} \label{eq_factorize}
\expvd{\hat{V}_{\alpha,L}(\tau) \hat{V}_{\beta,R}^{\dagger}(t)} & = & \expvd{\hat{V}_{\alpha,L}(\tau)} \expvd{\hat{V}_{\beta,R}^{\dagger}(t)} \nonumber \\
& = & \expvd{\hat{V}_{L}(\tau)} \expvd{\hat{V}_{L}^{\dagger}(t)} ~\mbox{.}
\end{eqnarray}
This gives rise to the coherent term in Eq. (\ref{eq_defcohincoh}),
\begin{eqnarray} \label{eq_sigcoh}
S_C(\omega_s) & = & \frac{2 \pi}{\Omega} \omega_s \left \vert P(\omega_s) \right \vert^2  ~\mbox{,} \nonumber \\ 
P(\omega_s) & = & \int_{-\infty}^{+\infty} \mathrm{d} \tau \mathrm{e}^{i \omega_s \tau} \expvd{\hat{V}_L^{\dagger}(\tau)} ~\mbox{.}
\end{eqnarray}
For this we have also made use of
\begin{equation}
\int_{-\infty}^{+\infty} \mathrm{d} t \int_{-\infty}^{t} \mathrm{d} \tau = \frac{1}{2} \int_{-\infty}^{+\infty} \mathrm{d} t \int_{-\infty}^{+ \infty} \mathrm{d} \tau ~\mbox{,}
\end{equation}
in Eq. (\ref{eq_sigincoh}).

In Eq. (\ref{eq_sigcoh}) we assumed that the sample is much smaller than the optical wavelength or that we have exact phase matching $\Delta \mathbf{k} = 0$, where $\Delta \mathbf{k}$ is the difference between the wavevector of the signal mode and the sum of wavevectors of the $n$ incoming fields. More generally, we should replace the $N(N-1)$ factor in Eq. (\ref{eq_defcohincoh}) by
\begin{equation} \label{eq_cohnwm_3}
F(\Delta \mathbf{k}) = \sum_{\alpha,\beta} \mathrm{e}^{-i \Delta \mathbf{k} \cdot \left( \mathbf{r}_{\alpha} - \mathbf{r}_{\beta} \right)} ~\mbox{.}
\end{equation}
For macroscopic samples, the number of terms with $\alpha \neq \beta$ in the double sum of Eq. (\ref{eq_cohnwm_3}) will by far exceed the terms with equal indices. Evaluating this double sum in the continuous limit and letting the sample volume go to infinity, gives the standard phase matching condition $F(\Delta \mathbf{k}) \rightarrow \delta(\Delta\mathbf{k})$, which yields a directed signal.

For small samples both coherent and incoherent terms need to be considered. It is interesting to note that for a single molecule we never get $S_C$, we only have $S_I$. {\em{Nonlinear susceptibilities, even though they are calculated for single molecules, do not represent a spontaneous wave-mixing experiment performed on a single molecule.}} Using the semiclassical approach one may conclude that coherent signals are possible even from a single molecule. The QED approach shows that this is not possible.

Below we compare the expressions for $S_I^{(n)}$ and $P^{(n)}$ obtained by expanding Eqs. (\ref{eq_sigincoh}) and (\ref{eq_sigcoh}) for $n=1,2,3$, where $n$ denotes the number of incoming modes. We demonstrate that the incoherent signal for $n$ incoming modes, relates to a $2(n+1)$-point correlation function in the dipole moment, compared to an $(n+1)$-point correlation function contained in $P^{(n)}$.

For $n=1$
\begin{eqnarray}
& P^{(1)}(\omega_s)=\int \mathrm{d}t \mathrm{e}^{i \omega_s t} \int_{-\infty}^{t} \mathrm{d} \tau \expv{\mathcal{T} \hat{V}_L(t) \hat{V}_{-}^{\dagger}(\tau)} \mathcal{E}(\mathrm{r},\tau) ~\mbox{,} \nonumber
\end{eqnarray}
gives Rayleigh scattering, whereas
\begin{eqnarray}
& S_I^{(1)}(\omega_s)=\int \mathrm{d}t \int_{-\infty}^{t} \mathrm{d} \tau \mathrm{e}^{i \omega_s (t-\tau)} \int \mathrm{d} \tau_1 \int \mathrm{d} \tau_2 \expv{\mathcal{T} \hat{V}_L(t) \hat{V}_R^{\dagger}(\tau) \hat{V}_{-}(\tau_1) \hat{V}_{-}^{\dagger}(\tau_2)} \mathcal{E}^{*}(\mathbf{r},\tau_1) \mathcal{E}(\mathbf{r},\tau_2) \nonumber
\end{eqnarray}
is responsible for SLE (Raman and Fluorescence) as discussed in Sec. \ref{sec_sle}.

For $n=2$
\begin{eqnarray}
& P^{(2)}(\omega_s)=\int \mathrm{d}t \mathrm{e}^{i \omega_s t} \int_{-\infty}^{t} \mathrm{d} \tau \int \mathrm{d} \tau_1 \expv{\mathcal{T} \hat{V}_L(t) \hat{V}_{-}^{\dagger}(\tau) \hat{V}_{-}(\tau_1)} \mathcal{E}(\mathrm{r},\tau) \mathcal{E}^{*}(\mathbf{r},\tau_1) \nonumber
\end{eqnarray}
describes 3-wave mixing e.g. second harmonic generation \cite{SHG_1}-\cite{SHG_3}, and
\begin{eqnarray}
&  S_I^{(2)}(\omega_s)=\int \mathrm{d}t \int_{-\infty}^{t} \mathrm{d} \tau \mathrm{e}^{i \omega_s (t-\tau)} \int \mathrm{d} \tau_1 \int \mathrm{d} \tau_2 \int \mathrm{d} \tau_3 \int \mathrm{d} \tau_4 \mathcal{E}^{*}(\mathbf{r},\tau_1) \mathcal{E}(\mathbf{r},\tau_2)  \mathcal{E}^{*}(\mathbf{r},\tau_3) \mathcal{E}(\mathbf{r},\tau_4) \nonumber \\
& \expv{\mathcal{T} \hat{V}_L(t) \hat{V}_R^{\dagger}(\tau) \hat{V}_{-}(\tau_1) \hat{V}_{-}^{\dagger}(\tau_2) \hat{V}_{-}(\tau_3) \hat{V}_{-}^{\dagger}(\tau_4)} \nonumber
\end{eqnarray}
may represent many possible processes, depending on the level scheme and off-resonant detunings, e.g. two photon fluorescence \cite{TPF_1}-\cite{TPF_3}.

For $n=3$
\begin{eqnarray}
& P^{(3)}(\omega_s)=\int \mathrm{d}t \mathrm{e}^{i \omega_s t} \int_{-\infty}^{t} \mathrm{d} \tau \int \mathrm{d} \tau_1 \int \mathrm{d} \tau_2 \mathcal{E}(\mathrm{r},\tau) \mathcal{E}(\mathbf{r},\tau_1) \mathcal{E}^{*}(\mathbf{r},\tau_1) \nonumber \\
& \expv{\mathcal{T} \hat{V}_L(t) \hat{V}_{-}^{\dagger}(\tau) \hat{V}_{-}(\tau_1) \hat{V}_{-}^{\dagger}(\tau_2)} \nonumber 
\end{eqnarray}
represents 4-wave mixing e.g. third harmonic generation and
\begin{eqnarray}
&  S_I^{(3)}(\omega_s)=\int \mathrm{d}t \int_{-\infty}^{t} \mathrm{d} \tau \mathrm{e}^{i \omega_s (t-\tau)} \int \mathrm{d} \tau_1 \int \mathrm{d} \tau_2 \int \mathrm{d} \tau_3 \int \mathrm{d} \tau_4 \int \mathrm{d} \tau_5 \int \mathrm{d} \tau_6 \mathcal{E}^{*}(\mathbf{r},\tau_1) \mathcal{E}(\mathbf{r},\tau_2)  \mathcal{E}^{*}(\mathbf{r},\tau_3) \nonumber \\ 
& \mathcal{E}(\mathbf{r},\tau_4) \mathcal{E}^{*}(\mathbf{r},\tau_5) \mathcal{E}(\mathbf{r},\tau_6) \expv{\mathcal{T} \hat{V}_L(t) \hat{V}_R^{\dagger}(\tau) \hat{V}_{-}(\tau_1) \hat{V}_{-}^{\dagger}(\tau_2) \hat{V}_{-}(\tau_3) \hat{V}_{-}^{\dagger}(\tau_4) \hat{V}_{-}(\tau_4) \hat{V}_{-}^{\dagger}(\tau_6)} \nonumber
\end{eqnarray}
can represent many possible processes, e.g. three photon fluorescence.

A nanocrystal is made out of many unit cells. Its dipole operator is given by $\hat{\mu}=\sum_{\alpha} \hat{\mu}^{\alpha}$, where $\hat{\mu}^{\alpha}$ represents an electron-hole pair on unit $\alpha$. It thus behaves as a collection of many molecules and it can show spontaneous coherent response \cite{Orrit1}-\cite{Orrit2}. $N$ is thus large even though we have a single particle.

\section{Heterodyne-detection as a stimulated wave mixing} \label{sec_het}

When detecting nonlinear optical signals in macroscopic samples, one can either measure the intensity (homodyne detection) or the magnitude and phase of the signal field (heterodyne detection). The later technique uses an intense external field $\mathcal{E}_{LO}$, usually referred to as ``local oscillator'', which interferes with the created nonlinear signal in the same direction, but is assumed not to interact with the molecule. Using the standard semiclassical approach to nonlinear optics, the quantity detected in a heterodyne measurement is \cite{mukbook}
\begin{equation} \label{eq_het}
S_{HET}(\mathbf{k}_s) \sim \int_{-\infty}^{+\infty} \mathrm{d} t \mathrm{Im} \lbrace \mathcal{E}_{LO}^{*}(t) P_s(\mathbf{k}_s,t) \rbrace ~\mbox{,}
\end{equation}
where $P_s$ is the polarization in the signal direction $\mathbf{k}_s$. With the QED formalism developed here, we can greatly simplify the derivation of Eq. (\ref{eq_het}) and show that it emerges naturally as an incoherent rather than a coherent signal. 

Equation (\ref{eq_cohnwm_1}) describes the signal generated from an assembly of particles by spontaneous emission. Applying Eq. (\ref{eq_macrdipole}) to Eq. (\ref{signal-7a}), we can give a similar expression for the stimulated signal
\begin{eqnarray} \label{eq_stim_assembl}
& S_{\mathrm{stim}}(t)= -\frac{2}{\hbar} \mbox{Re} \int_{-\infty}^t \mathrm{d} \tau
\left[\expvd{\mathcal{T} \hat{V}_{\alpha,R}(\tau) \hat{V}^{\dagger}_{\beta,L}(t)} - \expvd{\mathcal{T} \hat{V}_{\alpha,L}(\tau) \hat{V}^{\dagger}_{\beta,R}(t)} \right] \mathcal{E}(\br,t) \mathcal{E}^*(\br,\tau) &
\mbox{.} \nonumber\\
\end{eqnarray}
In analogy to Sec. (\ref{sec_cohincoh}), we can now split Eq. (\ref{eq_stim_assembl}) into a coherent ($\alpha \neq \beta$) and an incoherent ($\alpha = \beta$) contribution. Since the dipole operators of different, uncorrelated particles $\alpha \neq \beta$ commute, the coherent contribution vanishes identically. The present formalism hence yields the general result that stimulated processes are always incoherent in nature. In particular this applies to heterodyne experiments where the local oscillator serves as the stimulating field. It is hence sufficient to look only at a single particle and use Eq. (\ref{signal-2}) derived in Sec. \ref{sec_optsig}.

Equation (\ref{signal-2}) is reminiscent of the expression for heterodyne detection (Eq. (\ref{eq_het})), if one only looks at one particular mode ``s'' in the definition of the signal (see Eqs. (\ref{eq_defsig}) and (\ref{eq_defsig_1})), thus
\begin{eqnarray}
\label{eq_het_qed}
S_{\mathrm{SWM}}(t) = - \frac{2}{\hbar }\mbox{Im}\left[\expvdt{ \hat{\mathcal{E}}_s(\mathbf{r}_{\alpha},t)\hat{V}^{\dagger }}\right] ~\mbox{.}
\end{eqnarray}
Here, by adding a subscript $\alpha$, we make explicit that Eq. (\ref{eq_het_qed}) is the signal generated by {\em{one}} molecule at position $\mathbf{r}_{\alpha}$. For an assembly of noninteracting particles, this has then to be summed over $\mathbf{r}_{\alpha}$. The only difference of Eq. (\ref{eq_het_qed}) as compared to Eq. (\ref{eq_het}), is that it contains a field operator $\hat{\mathcal{E}}$, rather than a classical field $\mathcal{E}$. 

Heterodyne detection can be viewed as an incoherent stimulated emission process in the detected mode. Since Eq. (\ref{eq_het_qed}) is already first order in the mode $s$, in a subsequent perturbative expansion of the density operator the detecting mode will only contribute to higher order. Hence, to first order in the coupling with the detecting mode, the time-dependence of the density operator is only given by all other modes $s^{\prime} \neq s$. The field operator therfore acts directly on the state of the system (molecule + field) at $t \rightarrow -\infty$, which when assuming a coherent state of the field in the same limit (see Eq. (\ref{coherent-state})) yields
\begin{equation}
\label{eq_het_qed_2}
S(t)= \frac{2}{\hbar }\mbox{Im}\left[ \mathcal{E}_s^{*}(\br_{\alpha},t) \expvd{\hat{V}}\right] ~\mbox{.}
\end{equation}
In Eq. (\ref{eq_het_qed_2}) we additionally took the complex conjugate within the imaginary part.

The expectation value in Eq. (\ref{eq_het_qed_2}) can be further expanded in the incoming modes. For $n$ modes this gives the classical $n$th order polarization
\begin{equation}
P^{(n)}(\mathbf{r},t)=P^{(n)}(t)\mathrm{e}^{i \left(\sum_{s^{\prime} \neq s} \mathbf{k}_{s^{\prime}}\right) \cdot \mathbf{r}} ~\mbox{.}
\end{equation}
Consequently, integrating over all times, we obtain the following generic result for an $(n+1)$-wave mixing process (i.e. 1 detected mode ``s'' + $n$ incoming modes ``$s^{\prime} \neq s$'')
\begin{eqnarray}
\label{eq_nwm}
S(\Delta \mathbf{k}) & = & \frac{2}{\hbar}\int_{-\infty}^{+\infty} \mathrm{d} t \mbox{Im} \left[ \mathcal{E}_{s,0}^{*} P^{(n)}(t) f(\Delta \mathbf{k}) \right] ~\mbox{,} \nonumber \\
f(\Delta \mathbf{k}) & = & \sum_{\alpha} \mathrm{e}^{-i \Delta \mathbf{k} \cdot \mathbf{r}_{\alpha}} ~\mbox{,}
\end{eqnarray}
where we explicitly display the spatial dependence of the fields, e.g.
\begin{equation}
\mathcal{E}_s(\mathbf{r},t) =  \mathcal{E}_{s,0}(t) \mathrm{e}^{i \mathbf{k}_s \cdot \mathbf{r}} ~\mbox{.}
\end{equation}
In Eq. (\ref{eq_nwm}) we have added the sum over $\mathbf{r}_{\alpha}$ to make the transition from the signal generated by single molecule to one of several molecules (incoherent process).

Under exact phase-matching conditions ($\Delta \mathbf{k}=0$), or if the sample is much smaller than the optical wavelength, we have $f(\Delta \mathbf{k})=N$. Rather than creating a coherent signal, propagating it and interfering as done in the semiclassical approach, we can simply describe it microscopically as an incoherent signal.

Equation (\ref{eq_nwm}), hence allows us to interpret heterodyne detection as follows: {\em{All}} fields (the $n$ fields $(\omega_{s^{\prime}}, \mathbf{k}_{s^{\prime}})$ as well as the detected field $(\omega_{s}, \mathbf{k}_{s})$) interact with the molecule. This leads to stimulated emission of photons of frequency $\omega_s$. A sum over all molecular positions has to be performed (Eq. (\ref{eq_nwm})). Going to the continuum limit, which is justified for a macroscopic sample, this sum can be replaced by an integral over the sample volume. Extending this integral over the entire real space, we obtain the phase-matching condition $f(\Delta \mathbf{k}) \rightarrow \delta(\Delta \mathbf{k})$. Therefore, in a macroscopic sample the signal is finite only when the stimulating field $\mathcal{E}_s(\mathbf{r},t)$ fulfills this condition.

\section{conclusions} \label{sec_concl}

A microscopic QED treatment of nonlinear optical processes induced by a weak quantized field is developed in this paper. In the traditional, widely-used approaches to nonlinear optical response, the matter-field interaction is treated semiclassically, hence the signal field is obtained by simultaneously solving macroscopic (Maxwell-) and microscopic (quantum Liouville-) equations. Here we treat the entire process as a single event.

Based on a microscopic definition of the spontaneous and stimulated signal, general expressions are derived for the nonlinear optical signal involving non-equilibrium Green's functions in the incoming field. Depending on the experiment under consideration, these can then be expanded order by order in the incoming field modes. Application is made to spontaneous light emission and pump-probe spectroscopy. Coherent and incoherent processes involving both spontaneous and stimulated emission are treated using a unified framework.

A diagrammatic derivation of the contributing terms within the RWA using the technique of Keldysh Schwinger loop is developed. This respresentation reflects the partially time-ordered nature of the Green's functions and hence yields more compact expressions than the well established fully time ordered double sided Feynman diagrams. The loop diagrams are particularly useful for frequency-domain measurements where the bookeeping of time ordering is not necessary anymore. If needed for time-domain experiments, the fully time ordered expressions can be obtained from the loop diagrams using a simple prescription. Rules for constructing and reading these diagrams are laid out, making it possible to intuitively derive the expressions for the signals. 

The practical merits of this partially time-ordered approach are illustrated for the pump-probe technique where the number of diagrams is reduced from 16 to 8. This example also demonstrates an interesting fundamental merit of the QED treatment of nonlinear optical processes, even though in pump-probe spectroscopy quantum effects of the field (i.e. terms stemming from spontaneous emission) are usually negligible: Semiclassically, the pump-probe signal is obtained by calculating a third order response induced by two fields in the direction $\mathbf{k}_1 - \mathbf{k}_1 + \mathbf{k}_2$. The resulting signal field is then obtained using heterodyne detection, where mode 2 acts as its own local oscillator field. Even though the final expression for the signal is identical to a QED result, the semiclassical approach creats an artificial asymmetry between the local oscillator (mode 2), which by definition does not interact with the molecule, and the remaining fields (mode 1 and 2), which give rise to the nonlinear polarization. This asymmetry, which is eliminated here, is a direct consequence of the limitations imposed by separately treating matter and field within the semiclassical approach. 

Finally, using the QED formalism, we are able to treat both coherent and incoherent processes in a unified way. This distinction comes about by calculating the signal for an assembly of molecules rather than for a single particle. The signal then splits into two parts, a coherent term scaling $\sim N(N-1)$ and an incoherent term $\sim N$, where $N$ denotes the number of particles. As the number of particles in the sample is increased, the coherent term becomes more pronounced. For experiments carried out with only a few molecules however, both coherent and incoherent terms will generally make comparable contributions to the signal. In single molecule experiments only the incoherent term survives.

Since incoherent signals may not be calculated semiclassically, this leads to a striking consequence. For an 
$(n+1)$-wave mixing processes carried out on a single molecule we show that the signal is not given by an $(n+1)$-point response function of the dipole operator related to $P^{(n)}$, as predicted by the semiclassical approach. It is rather given by a different $2(n+1)$-point combination of correlation functions. This doubling arises since the signal may not be recast as an amplitude square and we must calculate the signal itself, not an amplitude. The present formalism allows a microscopic calculation of nonlinear optical experiments on single molecules which have only become feasible recently.

Finally, we address this apparent limitation of the semiclassical description from a more general viewpoint and show that heterodyne detection for an $(n+1)$-wave mixing experiment can be simply viewed as an incoherent stimulated emission process in the detected mode. In contrast, homodyne-detected $n$-wave mixing is a coherent spontaneous emission process. For a macroscopic sample, by summing over the molecules in the interaction volume, we recover the phase-matching condition, i.e. the heterodyne-signal is finite only if the stimulating mode (i.e. the local oscillator) matches this condition.

\appendix 
\section{Derivation of Eq. (\ref{eq_pp_chi3})} \label{app_chi3}
In this appendix we illustrate the equivalence of the QED expression for the pump-probe signal with the classical signal expressed in terms of $\chi^{(3)}$. Starting from Eq. (\ref{pp-signal-2}) and applying the definition of Eq. (\ref{definition+-}), we can write
\begin{eqnarray}
\label{hint-+-} 
{\cal H}_{int-}(\tau)&=& \frac{1}{\sqrt{2}}
\left[\hat{\mathcal{E}}_+(\br,\tau)\hat{V}^\dag_-+\hat{\mathcal{E}}_-(\br,\tau)\hat{V}^\dag_+\right.\nonumber\\
&+&\left.\hat{\mathcal{E}}^\dag_+(\br,\tau)\hat{V}_-+\hat{\mathcal{E}}^\dag_-(\br,\tau)\hat{V}_+\right] ~\mbox{.}
\end{eqnarray}

Making use of Eq. (\ref{hint-+-}), the correlation function in Eq. (\ref{pp-signal-2}) can be recast in terms of correlations of the system and the field. Calculating the field correlations, only 6 of the possible 16 terms are nonzero, giving
\begin{eqnarray}
\label{signal-9} 
& 2 \expv{\mathcal{T}\hat{V}_{+}^{\dagger}(t)\hat{V}_{-}(\tau){\cal H^{\prime}}_{int-}(\tau_1){\cal H^{\prime}}_{int-}(\tau_2)}=\nonumber\\
&\expv{\mathcal{T}\hat{V}_{+}^{\dagger}(t)\hat{V}_{-}(\tau) \hat{V}_-^\dag(\tau_1)\hat{V}_-(\tau_2)} \expv{\mathcal{T}\hat{\mathcal{E}}_+(\br,\tau_1)\hat{\mathcal{E}}_+^{\dagger}(\br,\tau_2)} \nonumber\\
+&\expv{\mathcal{T} \hat{V}_{+}^{\dagger}(t)\hat{V}_{-}(\tau)
\hat{V}_-(\tau_1)\hat{V}_-^{\dagger}(\tau_2)} \expv{\mathcal{T}\hat{\mathcal{E}}_+^{\dagger}(\br,\tau_1)\hat{\mathcal{E}}_+(\br,\tau_2)} \nonumber\\
+&\expv{\mathcal{T} \hat{V}_{+}^{\dagger}(t)\hat{V}_{-}(\tau)
\hat{V}_-^\dag(\tau_1)\hat{V}_+(\tau_2)} \expv{\mathcal{T} \hat{\mathcal{E}}_+(\br,\tau_1)\hat{\mathcal{E}}_-^{\dagger}(\br,\tau_2)} \nonumber\\
+&\expv{\mathcal{T} \hat{V}_{+}^{\dagger}(t)\hat{V}_{-}(\tau)
\hat{V}_+(\tau_1)\hat{V}_-^{\dagger}(\tau_2)} \expv{\mathcal{T} \hat{\mathcal{E}}_-^{\dagger}(\br,\tau_1)\hat{\mathcal{E}}_+(\br,\tau_2)} \nonumber \\
+&\expv{\mathcal{T} \hat{V}_{+}^{\dagger}(t)\hat{V}_{-}(\tau)
\hat{V}_{-}(\tau_1) \hat{V}_{+}^{\dagger}(\tau_2)} \expv{\mathcal{T} \hat{\mathcal{E}}_{+}^{\dagger}(\mathbf{r},\tau_1) \hat{\mathcal{E}}_{-}(\mathbf{r},\tau_2)} \nonumber \\
+&\expv{\mathcal{T} \hat{V}_{+}^{\dagger}(t)\hat{V}_{-}(\tau)
\hat{V}_{+}^{\dagger}(\tau_1) \hat{V}_{-}(\tau_2)} \expv{\mathcal{T} \hat{\mathcal{E}}_{-}(\mathbf{r},\tau_1) \hat{\mathcal{E}}_{+}^{\dagger}(\mathbf{r},\tau_2)} ~\mbox{,} \nonumber \\
\end{eqnarray}
where the system is assumed to be in its ground state for $t\rightarrow-\infty$, hence \begin{equation}
\expv{\mathcal{T} \hat{V}_{+}^{\dagger}(t) \hat{V}_{-}(\tau) \hat{V}_{-}(\tau_1) \hat{V}_{-}(\tau_2)} = 0 ~\mbox{.}
\end{equation}

Note that since Eq. (\ref{pp-signal-2}) is symmetric with respect to the variables $\tau_1$ and $\tau_2$, Eq. (\ref{signal-9}) simplifies further and only includes three distinct terms upon integration, i.e.
\begin{eqnarray}
\label{signal-9a} 
& \int \mathrm{d}\tau_1 \int \mathrm{d}\tau_2 2 \expv{ \mathcal{T} \hat{V}_R(\tau)\hat{V}_L^\dag(t){\cal H^{\prime}}_{int-}(\tau_1){\cal H^{\prime}}_{int-}(\tau_2)}=\nonumber\\
& 2 \int \mathrm{d}\tau_1 \int \mathrm{d}\tau_2 \times \nonumber \\
&\expv{\mathcal{T}\hat{V}_{+}^{\dagger}(t)\hat{V}_{-}(\tau) \hat{V}_-^\dag(\tau_1)\hat{V}_-(\tau_2)} \expv{\mathcal{T} \hat{\mathcal{E}}_+(\br,\tau_1)\hat{\mathcal{E}}_+^{\dagger}(\br,\tau_2)} \nonumber\\
+&\expv{\mathcal{T}\hat{V}_{+}^{\dagger}(t)\hat{V}_{-}(\tau)
\hat{V}_-^\dag(\tau_1)\hat{V}_+(\tau_2)} \expv{\mathcal{T}\hat{\mathcal{E}}_+(\br,\tau_1)\hat{\mathcal{E}}_-^{\dagger}(\br,\tau_2)} \nonumber\\
+&\expv{\mathcal{T}\hat{V}_{+}^{\dagger}(t)\hat{V}_{-}(\tau)
\hat{V}_{-}(\tau_1) \hat{V}_{+}^{\dagger}(\tau_2)} \expv{\mathcal{T} \hat{\mathcal{E}}_{+}^{\dagger}(\mathbf{r},\tau_1) \hat{\mathcal{E}}_{-}(\mathbf{r},\tau_2)} ~\mbox{.} \nonumber \\
\end{eqnarray}
For a classical incoming field, the correlation function in Eq. (\ref{signal-9}) is equal to only the first term on the rhs, where
\begin{eqnarray} \label{eq_chi3_classterm}
\expv{\mathcal{T} \hat{\mathcal{E}}_{+}(\mathbf{r},\tau_1) \hat{\mathcal{E}}_{+}^{\dagger}(\mathbf{r},\tau_2)} = \mathcal{E}(\mathbf{r},\tau_1) \mathcal{E}^{*}(\mathbf{r},\tau_2) + \dfrac{2 \pi \hbar \omega}{\Omega} \mathrm{e}^{i \omega (\tau_1-\tau_2)} ~\mbox{.}
\end{eqnarray}
For a classical field the second term in Eq. (\ref{eq_chi3_classterm}) will be negligible.

The last two terms in Eq. (\ref{signal-9a}) are due to the quantum character of the incoming field, since 
\begin{eqnarray} \label{eq_chi3_neglterm}
\expv{\mathcal{T} \hat{\mathcal{E}}_{+}(\mathbf{r},\tau_1) \hat{\mathcal{E}}_{-}^{\dagger}(\mathbf{r},\tau_2)} = \dfrac{2 \pi \hbar \omega}{\Omega} \mathrm{e}^{-i \omega (\tau_1-\tau_2)} \theta(\tau_1-\tau_2) ~\mbox{,} \nonumber \\
\expv{\mathcal{T} \hat{\mathcal{E}}_{+}^{\dagger}(\mathbf{r},\tau_1) \hat{\mathcal{E}}_{-}(\mathbf{r},\tau_2)} = - \dfrac{2 \pi \hbar \omega}{\Omega} \mathrm{e}^{-i \omega (\tau_2-\tau_1)} \theta(\tau_1-\tau_2) ~\mbox{.}
\end{eqnarray}
Hence, neglecting the last two terms in Eq. (\ref{signal-9a}) and making use of Eq. (\ref{eq_chi3_classterm}), we obtain Eq. (\ref{eq_pp_chi3}).

\section{Time domain expressions for the pump-probe signal} \label{app_pp}

\begin{figure}[htbp]
\begin{center}
\subfigure[]{\includegraphics[width=0.2\columnwidth]{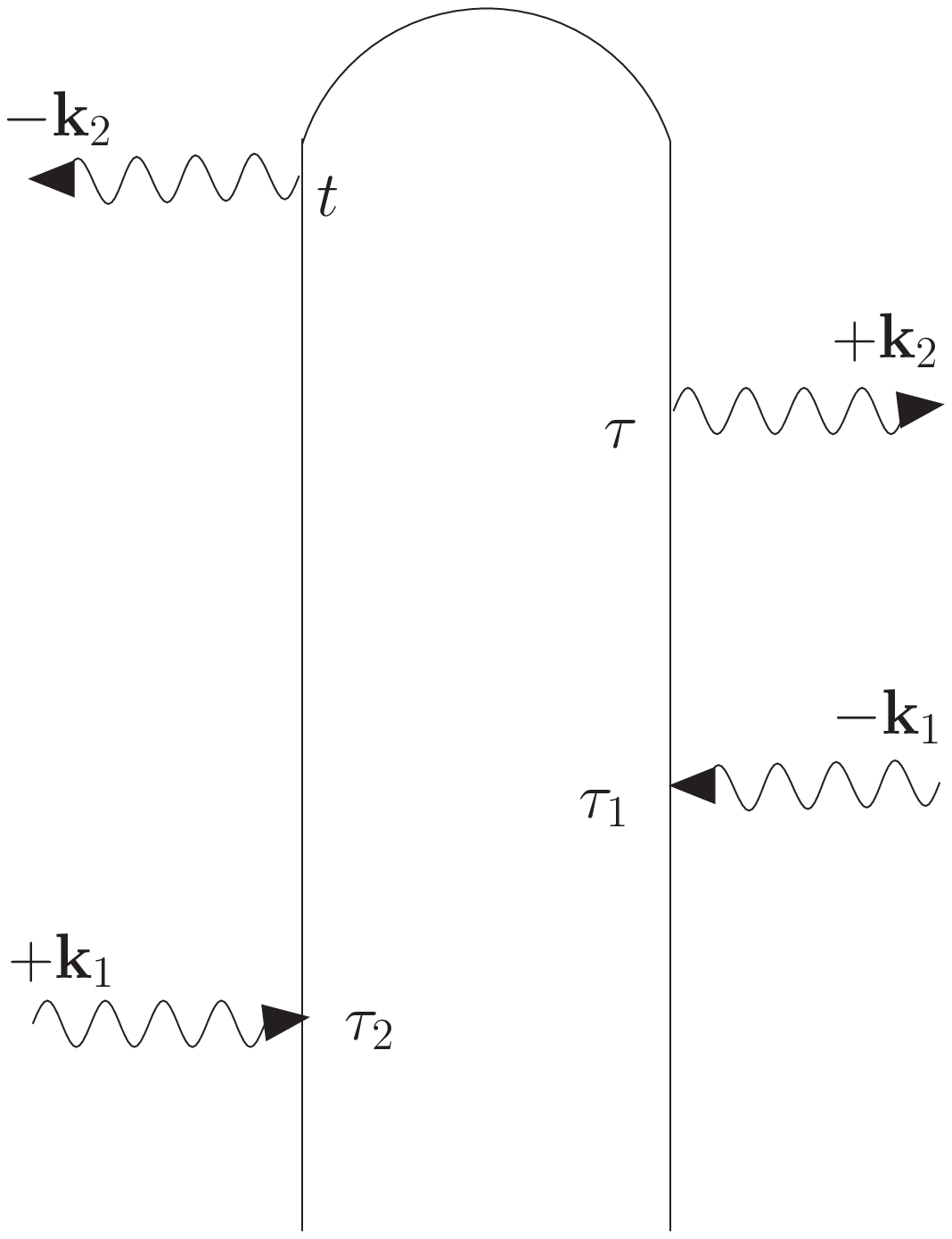}}
\subfigure[]{\includegraphics[width=0.2\columnwidth]{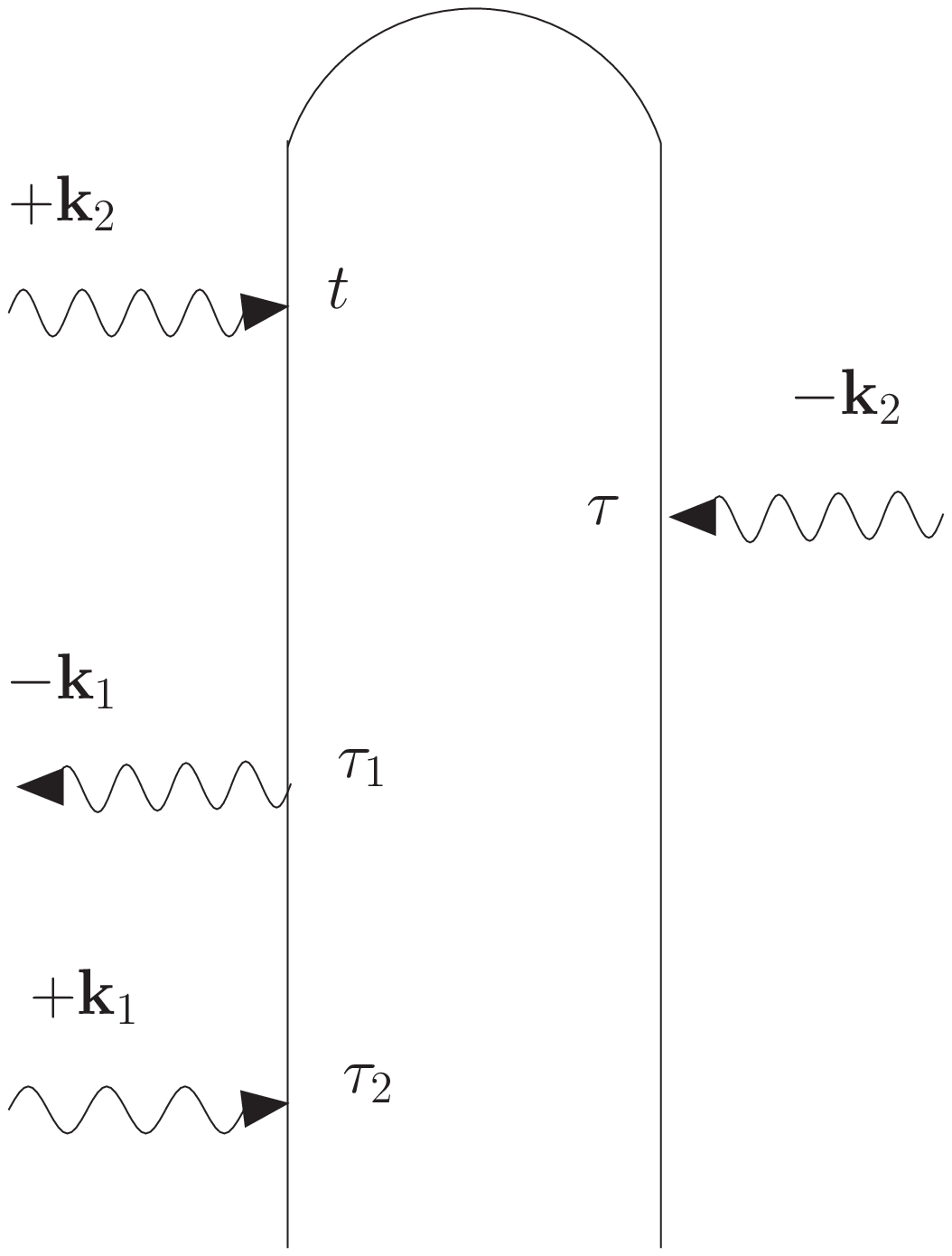}}
\subfigure[]{\includegraphics[width=0.2\columnwidth]{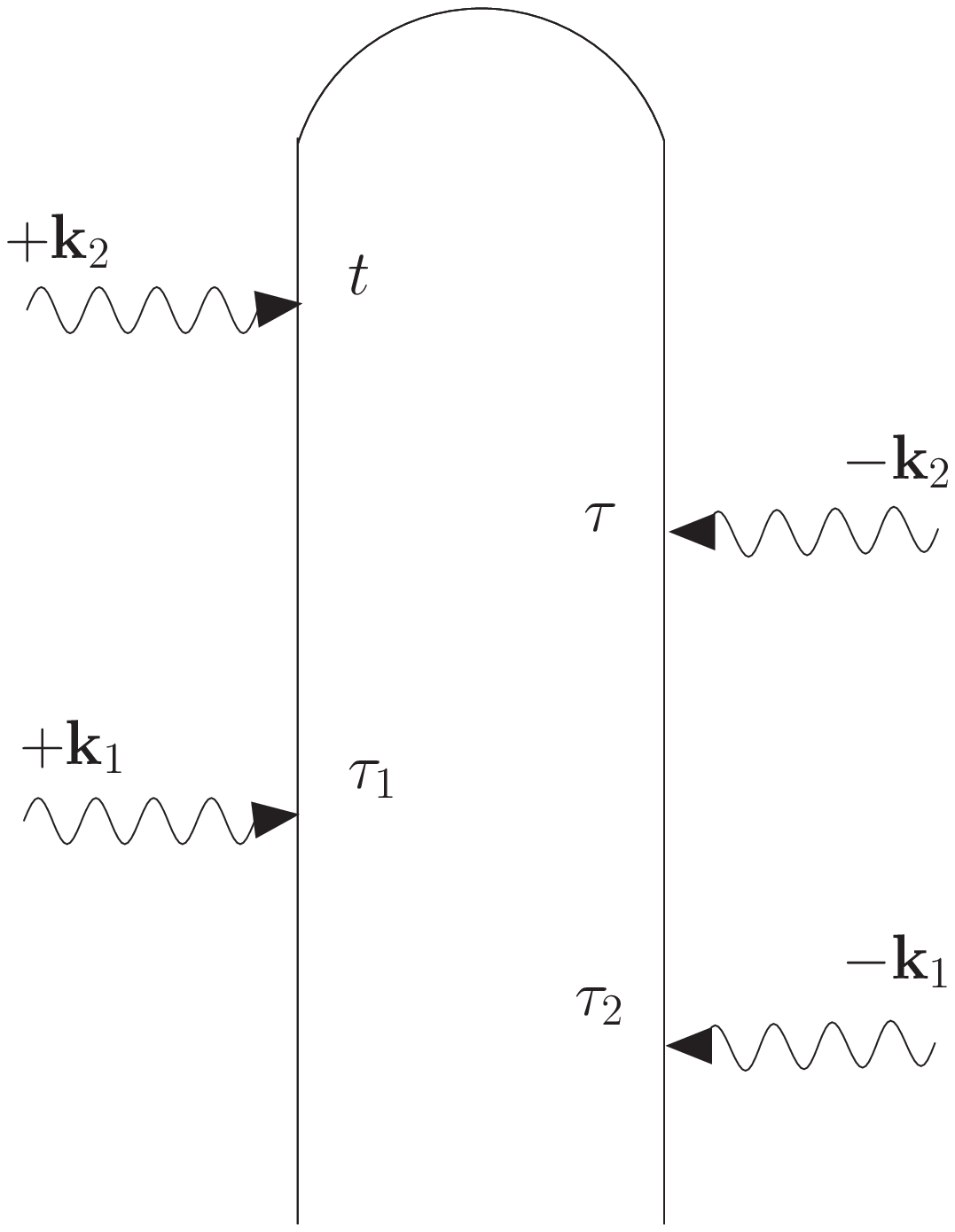}}
\subfigure[]{\includegraphics[width=0.2\columnwidth]{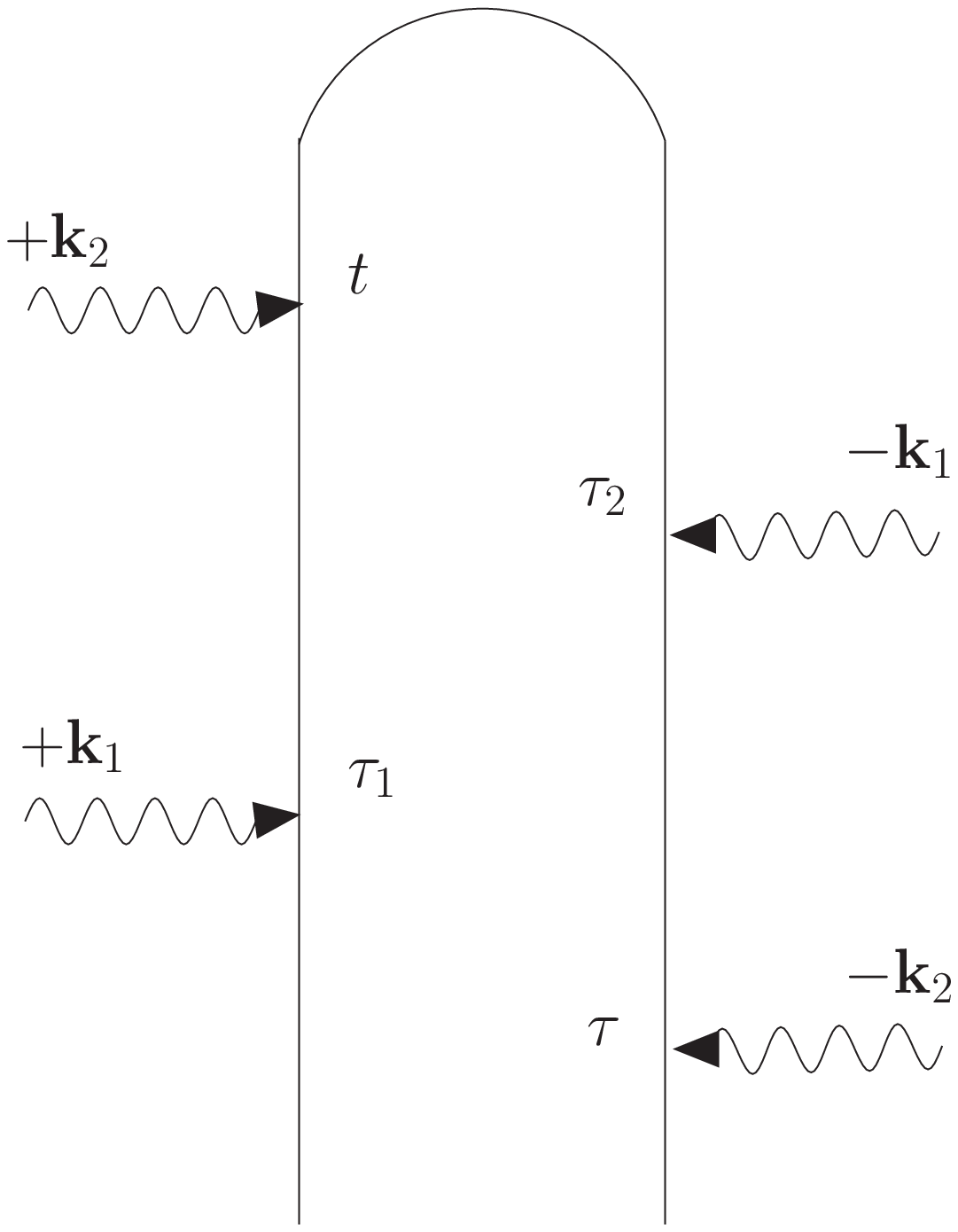}}
\subfigure[]{\includegraphics[width=0.2\columnwidth]{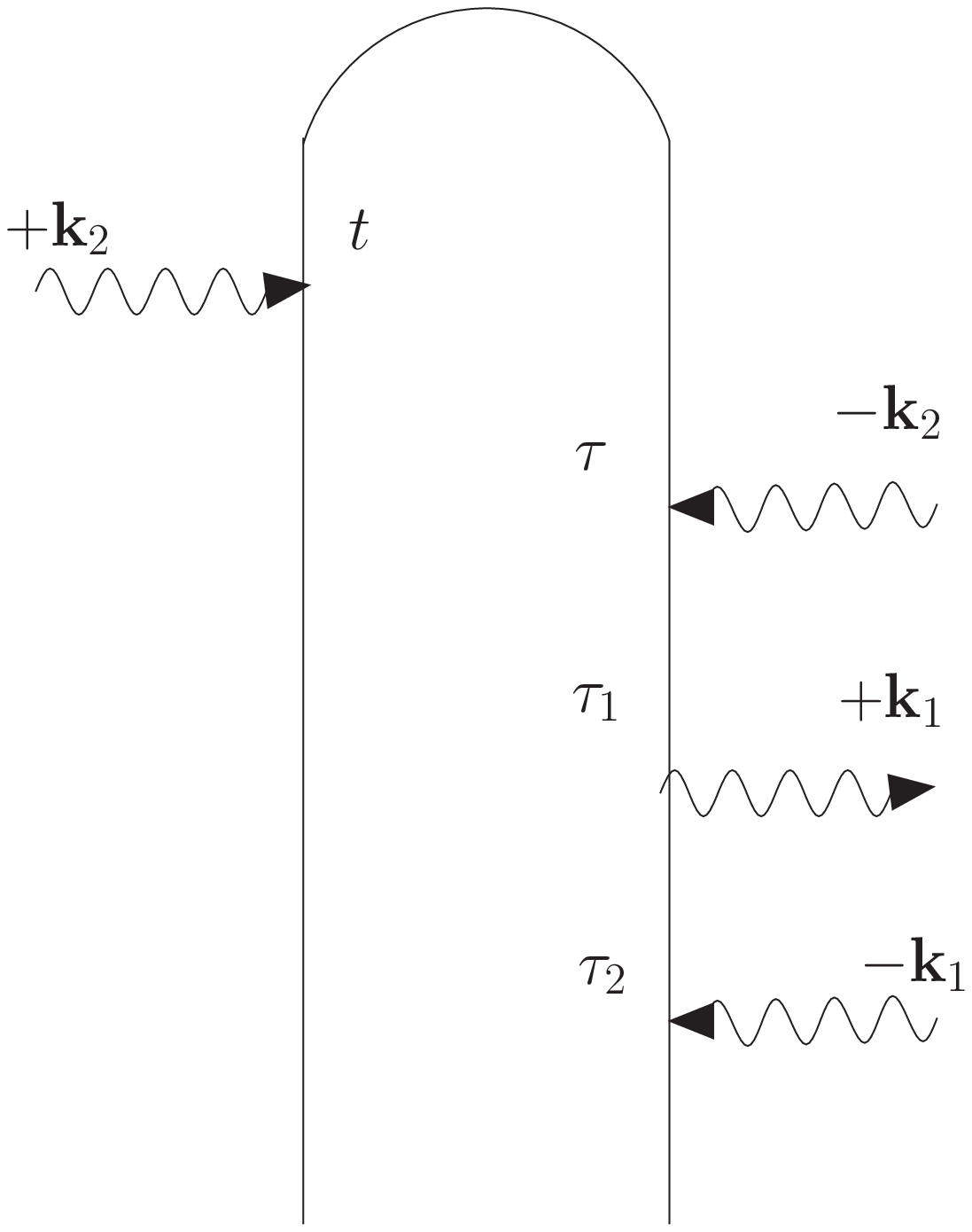}}
\subfigure[]{\includegraphics[width=0.2\columnwidth]{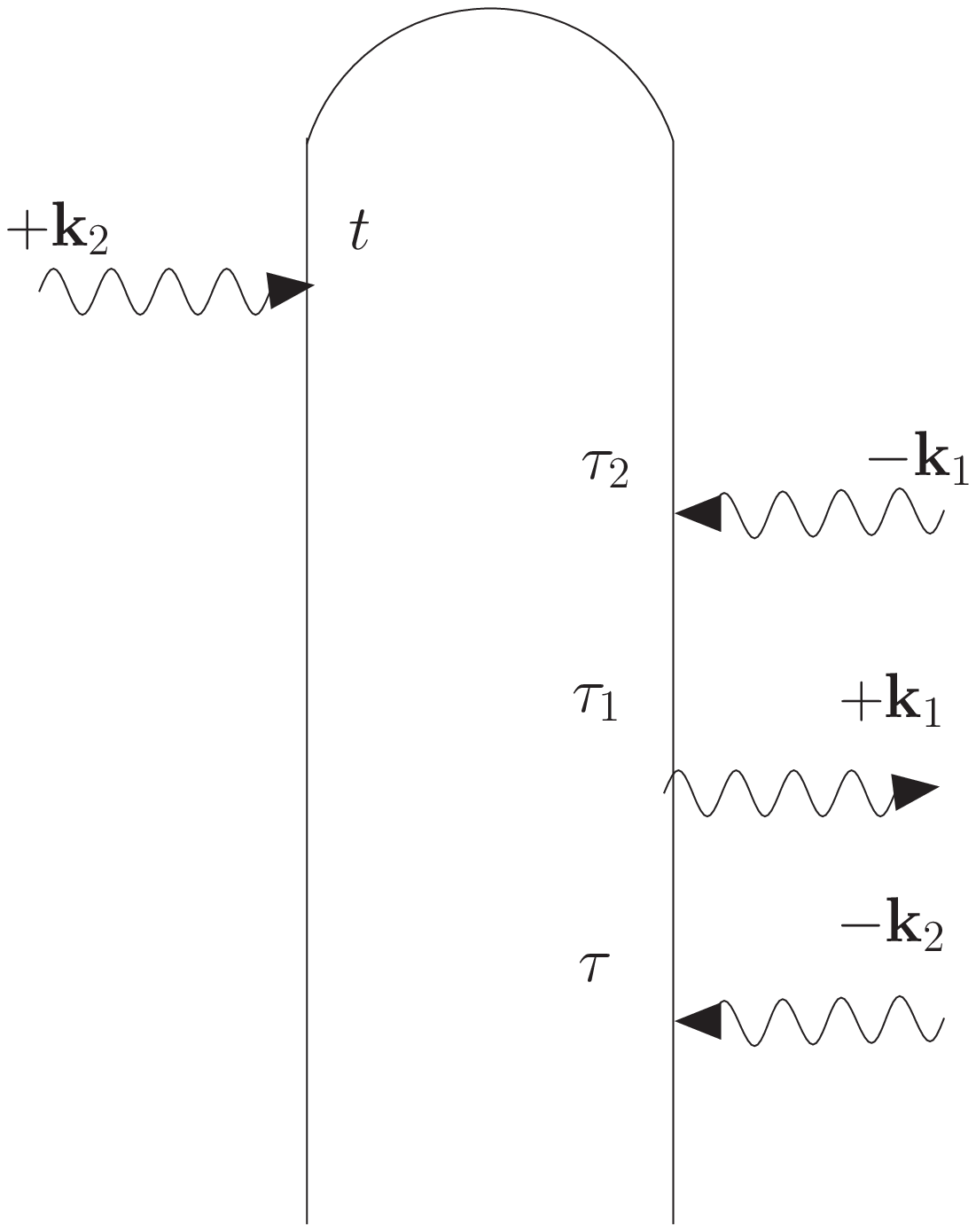}}
\subfigure[]{\includegraphics[width=0.2\columnwidth]{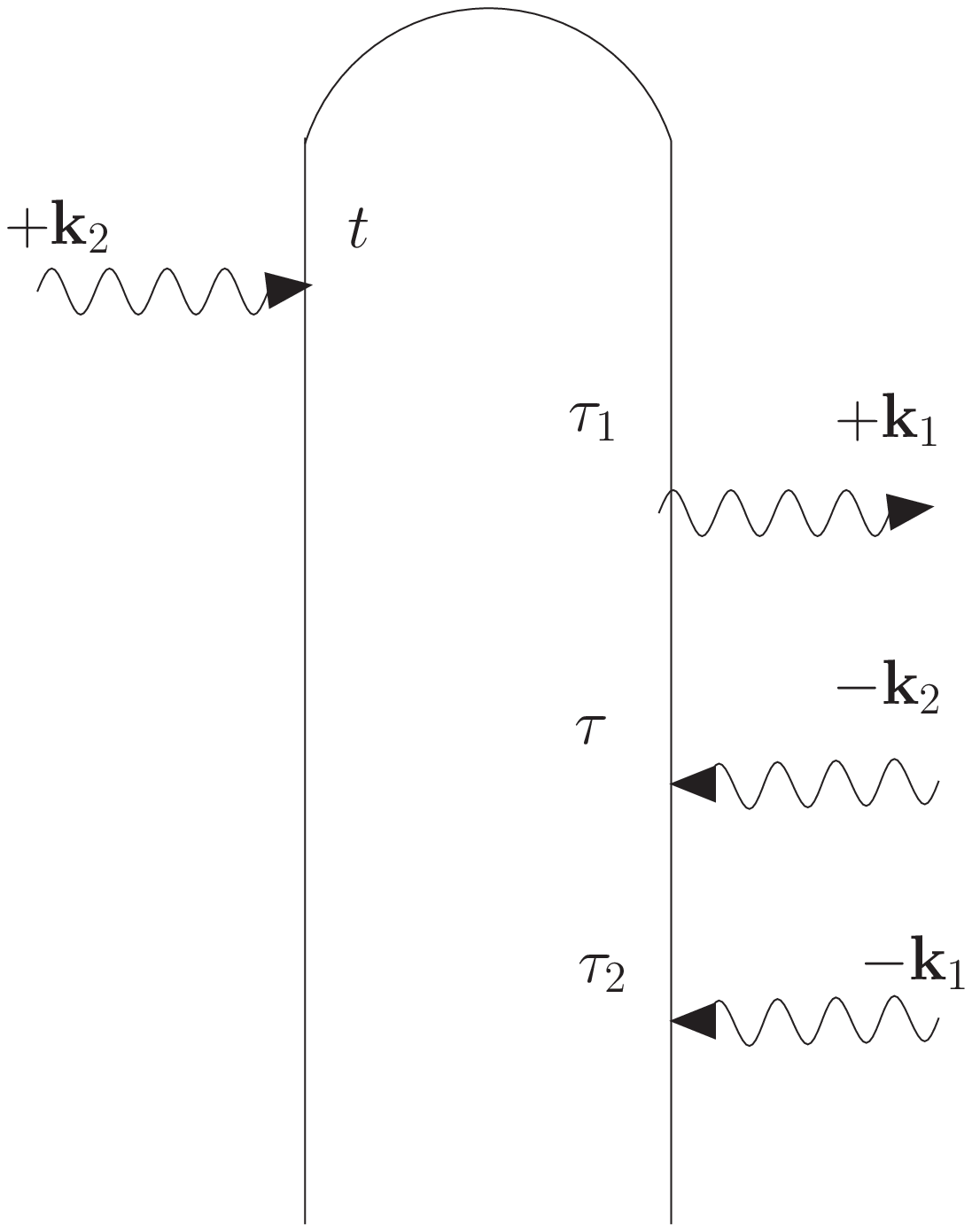}}
\subfigure[]{\includegraphics[width=0.2\columnwidth]{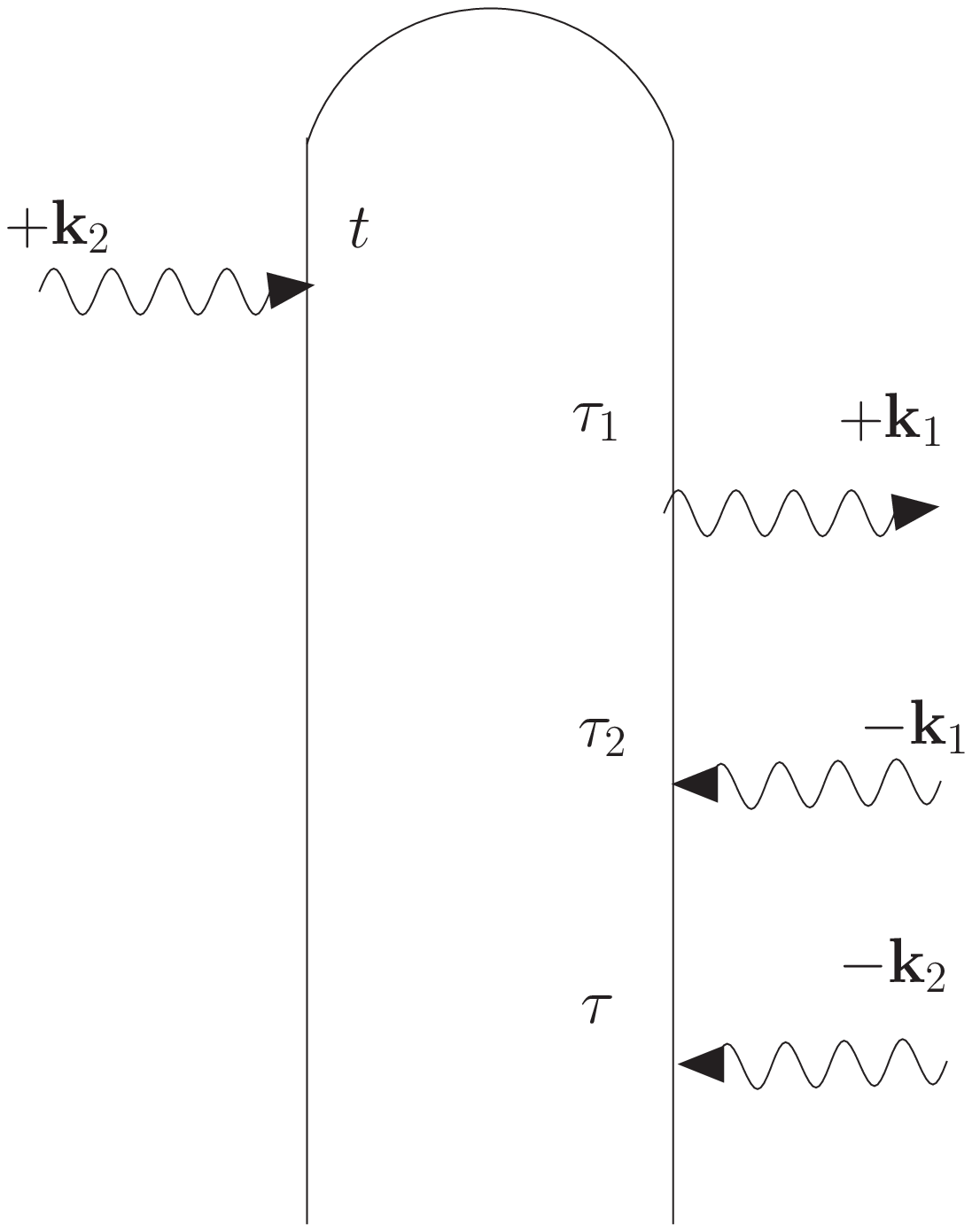}}
\caption{The eight loop diagrams for the pump-probe signal (Eq. (\ref{eq_pp_frequ})). Note that diagram (a) coincides with the SLE (Fig. \ref{fig_SLE_loop_1}).}
\label{fig_PP_loop}
\end{center}
\end{figure}

\begin{figure}[htbp]
\begin{center}
\subfigure[]{\includegraphics[width=0.6\columnwidth]{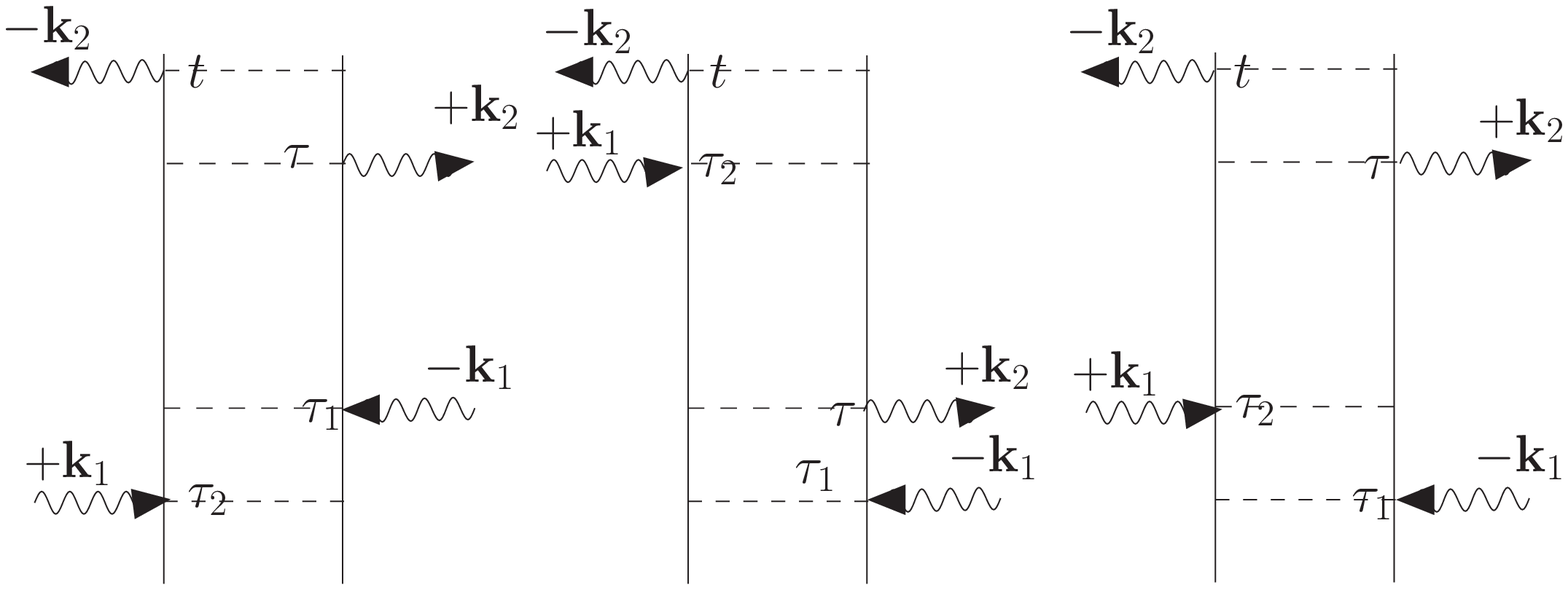}}
\subfigure[]{\includegraphics[width=0.6\columnwidth]{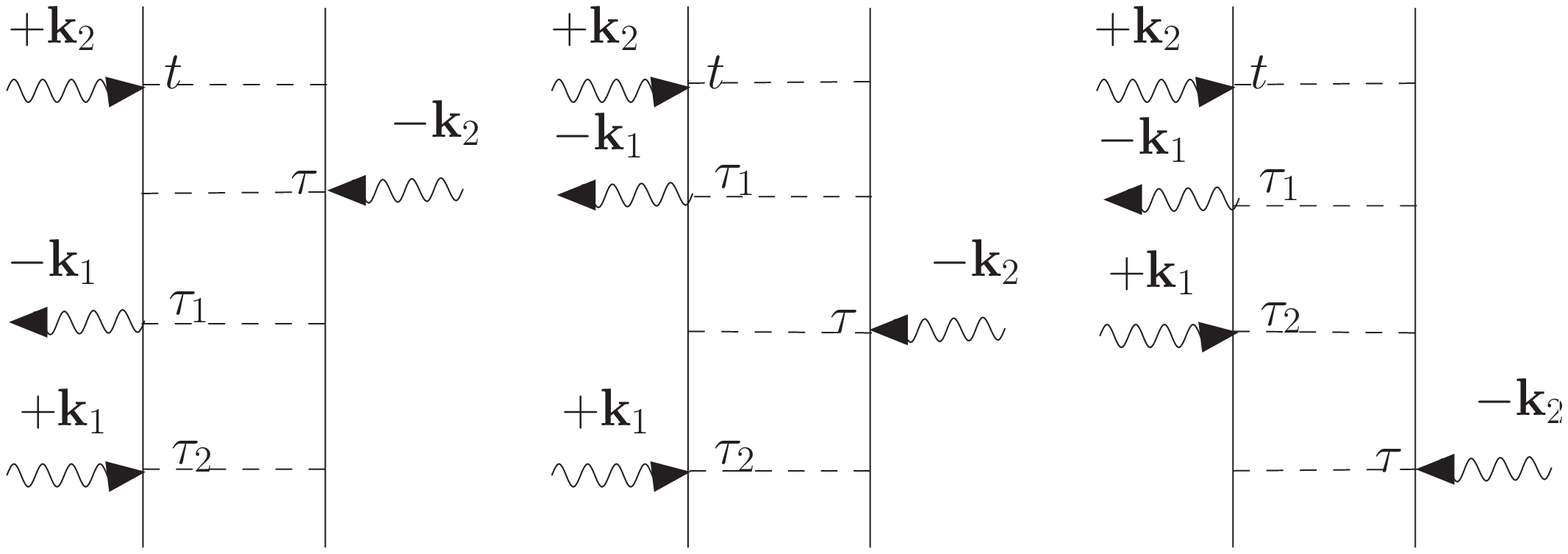}}
\subfigure[]{\includegraphics[width=0.6\columnwidth]{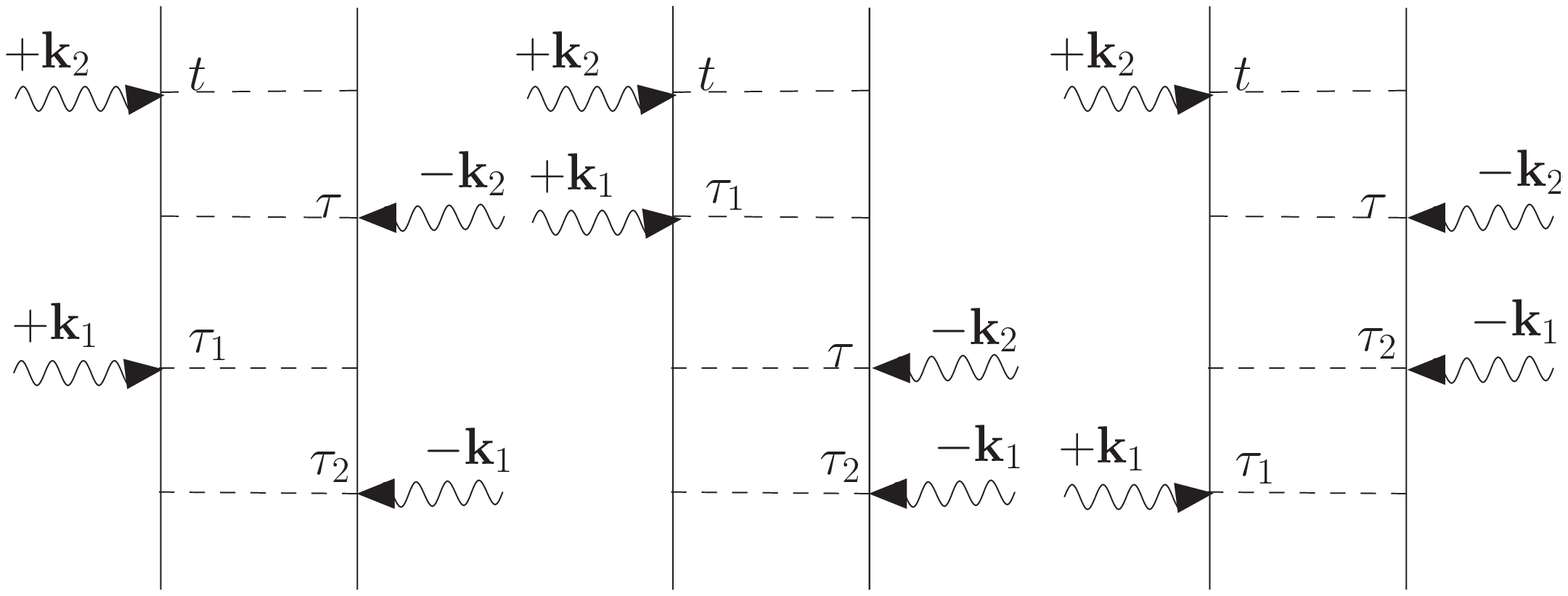}}
\subfigure[]{\includegraphics[width=0.6\columnwidth]{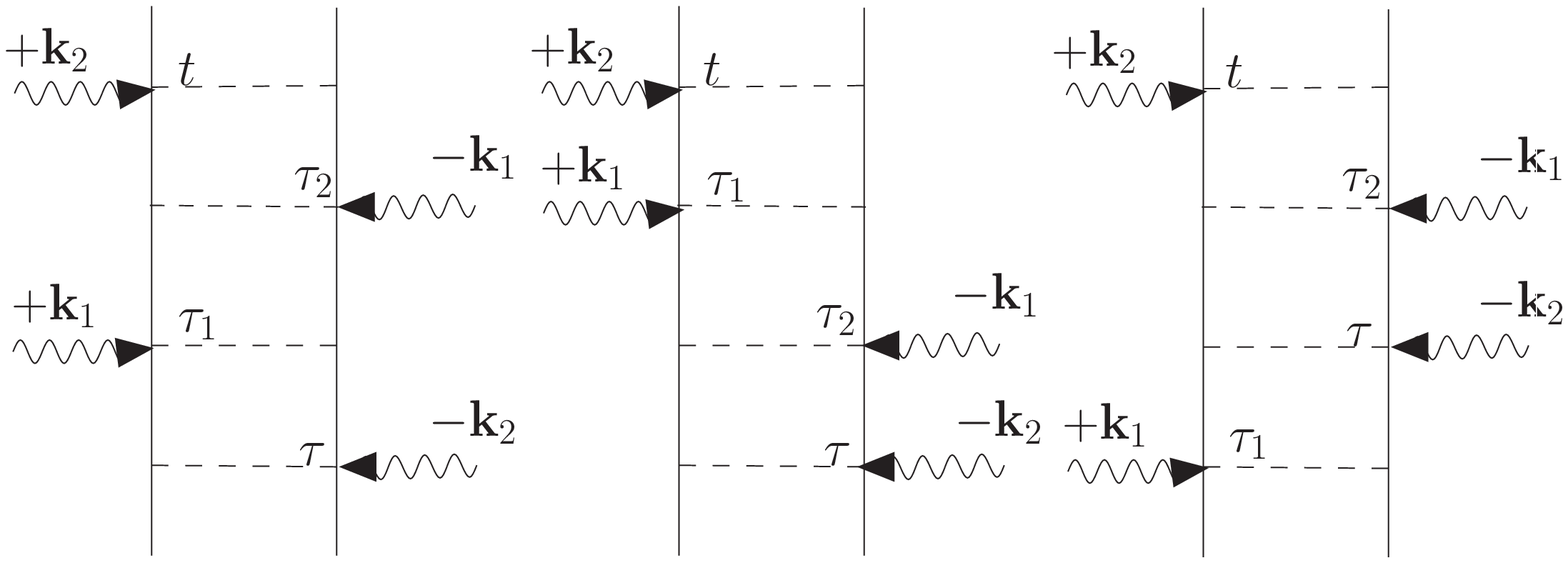}}
\caption{The Double sided Feynman diagrams resulting from Fig. \ref{fig_PP_loop} upon breaking the loops into time ordered contributions. Panel (a) - (d) are the time ordered diagrams corresponding to the loops shown in panel (a) - (d) of Fig. \ref{fig_PP_loop}, respectively. Since the four loops with three interactions on the bra (Fig. \ref{fig_PP_loop}, panels (e) - (h)) are already fully-time ordered, each gives a single double sided diagram. These are not repeated here. Overall, the eight loop diagrams of Fig. \ref{fig_PP_loop} yield 16 Feynman diagrams.}
\label{fig_PP_loop_open}
\end{center}
\end{figure}

In this appendix we give the expressions for the pump-probe signal in the time-domain obtained by applying the rules summarized in Sec. \ref{sec_sle} to the loop diagrams in Fig. \ref{fig_PP_loop}. This complements the frequency-domain expression given in Eq. (\ref{eq_pp_frequ}).
Applying the time-domain rules for the loop diagrams in Fig. \ref{fig_PP_loop}, we obtain
\begin{widetext}
\begin{eqnarray} \label{eq_pp-signal-2a}
& S_{\mathrm{PP}}(t) = \frac{4}{\hbar^3} \mathrm{Re} \left \lbrace \int_{-\infty}^{t} \mathrm{d}\tau \int \mathrm{d}\tau_1 \int \mathrm{d} \tau_2 \mathcal{E}_2^{*}(\mathbf{r},t)\mathcal{E}_2(\mathbf{r},\tau) \right.  \nonumber \\
& \expv{\mathcal{T} V_L(t) V_R^{\dagger}(\tau) V_L^{\dagger}(\tau_2) V_R(\tau_1)}  \expv{\mathcal{T}\hat{\mathcal{E}}_{1,L}(\mathbf{r},\tau_2) \hat{\mathcal{E}}_{1,R}^{\dagger}(\mathbf{r},\tau_1)}  \nonumber \\ 
& + \int_{-\infty}^{t} \mathrm{d}\tau \int \mathrm{d}\tau_1 \int \mathrm{d} \tau_2 \mathcal{E}_2(\mathbf{r},t)\mathcal{E}_2^{*}(\mathbf{r},\tau) \nonumber \\
& + \left[ \expv{\mathcal{T} V_L^{\dagger}(t) V_R(\tau) V_L(\tau_1) V_L^{\dagger}(\tau_2)} \expv{\mathcal{T} \hat{\mathcal{E}}_{1,L}^{\dagger}(\mathbf{r},\tau_1) \hat{\mathcal{E}}_{1,L}(\mathbf{r},\tau_2)} \right. \nonumber \\
& - \expv{\mathcal{T} V_L^{\dagger}(t) V_R(\tau) V_L^{\dagger}(\tau_1) V_R(\tau_2)} \expv{\mathcal{T} \hat{\mathcal{E}}_{1,L}(\mathbf{r},\tau_1) \hat{\mathcal{E}}_{1,R}^{\dagger}(\mathbf{r},\tau_2)} \nonumber \\
& \left. \left. + \expv{\mathcal{T} V_L^{\dagger}(t) V_R(\tau) V_R^{\dagger}(\tau_1) V_R(\tau_2)} \expv{\mathcal{T} \hat{\mathcal{E}}_{1,R}(\mathbf{r},\tau_1) \hat{\mathcal{E}}_{1,R}^{\dagger}(\mathbf{r},\tau_2)} \right] \right \rbrace ~\mbox{,} 
\end{eqnarray}
\end{widetext}
where we kept the order of the diagrams in Fig. \ref{fig_PP_loop}. Similar to Eq. (\ref{eq_spont_1}), an additional factor 2 accounts for the symmetry with respect to $\tau_1$ and $\tau_2$. This can again be broken into fully time-ordered terms using the double sided Feynman diagrams displayed in Fig. \ref{fig_PP_loop_open}. We then get
\begin{widetext}
\begin{eqnarray} \label{eq_pp-signal-2b}
& & S_{\mathrm{PP}}(t) = \frac{4}{\hbar^3} \mathrm{Re}  \int_{-\infty}^{t} \mathrm{d}\tau \int \mathrm{d}\tau_1 \int \mathrm{d} \tau_2 \mathcal{E}_2^{*}(\mathbf{r},t) \mathcal{E}_2(\mathbf{r},\tau) \nonumber \\
& & \mathcal{E}_1^{*}(\mathbf{r},\tau_1) \mathcal{E}_1(\mathbf{r},\tau_2) \left[ \theta(t \tau) \theta(\tau \tau_1) \theta(\tau_1 \tau_2) \expv{\hat{V}_L(t) \hat{V}_R^{\dagger}(\tau) \hat{V}_R(\tau_1) \hat{V}_L^{\dagger}(\tau_2)} \right. \nonumber \\
& + & \theta(t \tau_2) \theta(\tau_2 \tau) \theta(\tau \tau_1) \expv{\hat{V}_L(t) \hat{V}_L^{\dagger}(\tau_2) \hat{V}_R^{\dagger}(\tau) \hat{V}_R(\tau_1)} \nonumber \\
& + & \left. \theta(t \tau_2) \theta(\tau_2 \tau) \theta(\tau \tau_1) \expv{\hat{V}_L(t) \hat{V}_L^{\dagger}(\tau_2) \hat{V}_R^{\dagger}(\tau) \hat{V}_R(\tau_1)} + \theta(t \tau) \theta(\tau \tau_2) \theta(\tau_2 \tau_1) \expv{\hat{V}_L(t) \hat{V}_R^{\dagger}(\tau) \hat{V}_L^{\dagger}(\tau_2) \hat{V}_R(\tau_1)} \right] \nonumber \\
& + & \mathcal{E}_2(\mathbf{r},t) \mathcal{E}_2^{*}(\mathbf{r},\tau)
\left \lbrace \mathcal{E}_1^{*}(\mathbf{r},\tau_1) \mathcal{E}_1(\mathbf{r},\tau_2) \left[ \theta(t \tau) \theta(\tau \tau_1) \theta(\tau_1 \tau_2) \expv{\hat{V}_L^{\dagger}(t) \hat{V}_R(\tau) \hat{V}_L(\tau_1) \hat{V}_L^{\dagger}(\tau_2)}  \right. \right. \nonumber \\
& + & \left. \theta(t \tau_1) \theta(\tau_1 \tau) \theta(\tau \tau_2) \expv{\hat{V}_L^{\dagger}(t) \hat{V}_L(\tau_1) \hat{V}_R(\tau) \hat{V}_L^{\dagger}(\tau_2)} + \theta(t \tau_1) \theta(\tau_1 \tau_2) \theta(\tau_2 \tau) \expv{\hat{V}_L^{\dagger}(t) \hat{V}_L(\tau_1) \hat{V}_L^{\dagger}(\tau_2) \hat{V}_R(\tau)} \right] \nonumber \\
& - & \mathcal{E}_1(\mathbf{r},\tau_1) \mathcal{E}_1^{*}(\mathbf{r},\tau_2)) \left[ \theta(t \tau) \theta(\tau \tau_1) 
\theta(\tau_1 \tau_2) \expv{\hat{V}_L^{\dagger}(t) \hat{V}_R(\tau) \hat{V}_L^{\dagger}(\tau_1) \hat{V}_R(\tau_2)} \right.  \nonumber \\
& + & \theta(t \tau_1) \theta(\tau_1 \tau) \theta(\tau \tau_2) \expv{\hat{V}_L^{\dagger}(t) \hat{V}_L^{\dagger}(\tau_1) \hat{V}_R(\tau) \hat{V}_R(\tau_2)} + \theta(t \tau) \theta(\tau \tau_2) \theta(\tau_2 \tau_1) \expv{\hat{V}_L^{\dagger}(t) \hat{V}_R(\tau) \hat{V}_R(\tau_2) \hat{V}_L^{\dagger}(\tau_1)} \nonumber \\
& + & \theta(t \tau_2) \theta(\tau_2 \tau_1) \theta(\tau_1 \tau) \expv{\hat{V}_L^{\dagger}(t) \hat{V}_R^{\dagger}(\tau_2) \hat{V}_L^{\dagger}(\tau_1) \hat{V}_R(\tau)} + \theta(t \tau_1) \theta(\tau_1 \tau_2) \theta(\tau_2 \tau) \expv{\hat{V}_L^{\dagger}(t) \hat{V}_L^{\dagger}(\tau_1) \hat{V}_R(\tau_2) \hat{V}_R(\tau)} \nonumber \\
& + & \left. \theta(t \tau_2) \theta(\tau_2 \tau) \theta(\tau \tau_1) \expv{\hat{V}_L^{\dagger}(t) \hat{V}_R(\tau_2) \hat{V}_R(\tau) \hat{V}_L^{\dagger}(\tau_1)} \right] \nonumber \\ 
& + & \mathcal{E}_1(\mathbf{r},\tau_1) \mathcal{E}_1^{*}(\mathbf{r},\tau_2) \left[ \theta(t \tau) \theta(\tau \tau_1) \theta(\tau_1 \tau_2) \expv{\hat{V}_L^{\dagger}(t) \hat{V}_R(\tau) \hat{V}_R^{\dagger}(\tau_1) \hat{V}_R(\tau_2)}   \right. \nonumber \\
& + & \theta(t \tau_2) \theta(\tau_2 \tau_1) \theta(\tau_1 \tau) \expv{\hat{V}_L^{\dagger}(t) \hat{V}_R(\tau_2) \hat{V}_R^{\dagger}(\tau_1) \hat{V}_R(\tau)}
+ \theta(t \tau_1) \theta(\tau_1 \tau) \theta(\tau \tau_2) \expv{\hat{V}_L^{\dagger}(t) \hat{V}_R^{\dagger}(\tau_1) \hat{V}_R(\tau) \hat{V}_R(\tau_2)} \nonumber \\
& + & \left. \left. \theta(t \tau_1) \theta(\tau_1 \tau_2) \theta(\tau_2 \tau) \expv{\hat{V}_L^{\dagger}(t) \hat{V}_R^{\dagger}(\tau_1) \hat{V}_R(\tau_2) \hat{V}_R(\tau)} \right] \right \rbrace ~\mbox{.} 
\end{eqnarray}
\end{widetext}

Comparison of Eqs. (\ref{eq_pp-signal-2a}) and (\ref{eq_pp-signal-2b}) illustrates the tremendous simplification achieved by the more compact partially-time-ordered approach. Employing the diagrammatic Keldysh-Schwinger loop technique, the number of diagrams is reduced by half (16 vs. 8 diagrams).

\acknowledgements{We wish to thank Prof. Sunney Xie for useful discussions. The support of the National Science Foundation (Grant No. CHE-0446555, CBC-0533162), NIRT
(Grant No. EEC 0303389)) and the National Institutes of Health (GM59230) is gratefully acknowledged.}


\newpage

\newpage

\newpage

\bigskip

\end{document}